\documentclass[sigconf]{acmart}

\AtBeginDocument{%
  }

\usepackage{graphicx}
\usepackage{float}
\usepackage{booktabs}
\usepackage{tabularx}
\usepackage{longtable}
\usepackage{array}
\usepackage{pdflscape}
\usepackage{xurl}
\usepackage{placeins}
\usepackage{amsmath}
\usepackage{enumitem}
\usepackage[skins]{tcolorbox}
\definecolor{cognitiveblue}{HTML}{356288}

\graphicspath{{figures/}}
\newcolumntype{Y}{>{\raggedright\arraybackslash}X}
\newcolumntype{P}[1]{>{\raggedright\arraybackslash}p{#1}}
\definecolor{fillinyellow}{RGB}{255,242,153}

\setcopyright{none}
\usepackage{tikz}
\usetikzlibrary{arrows.meta}
\usepackage{xcolor}
\usepackage{soul}
\providecommand{\findingclaim}[1]{%
  \begingroup
    \sethlcolor{blue!8}%
    \textbf{\hl{#1}}%
  \endgroup
}

\usepackage{listings}
\lstdefinestyle{codingprompt}{%
  basicstyle=\scriptsize\ttfamily,
  breaklines=true,
  breakautoindent=true,
  breakindent=1.2em,
  columns=fullflexible,
  keepspaces=true,
  showstringspaces=false,
  frame=tb,
  framesep=5pt,
  rulecolor=\color{black!45},
  xleftmargin=0pt,
  xrightmargin=0pt,
  aboveskip=10pt,
  belowskip=10pt,
  captionpos=b,
  numbers=none,
  upquote=true,
  literate={-}{{-}}1
}

\begin{document}

\title[Reconceptualizing Age Assurance as a Sociotechnical Problem]{Reconceptualizing Age Assurance as a Sociotechnical Problem: Connecting Evidence, Evaluation, Claims, and Decisions}

\author{Renkai Ma}
\email{mark@ucmail.uc.edu}
\affiliation{%
  \department{School of Information Technology}
  \institution{University of Cincinnati}
  \city{Cincinnati}
  \state{Ohio}
  \country{United States}
}

\author{Prakriti Dumaru}
\email{pdumaru@icsi.berkeley.edu}
\affiliation{%
  \department{Socio-Technical Interaction Research Lab}
  \institution{International Computer Science Institute}
  \city{Berkeley}
  \state{California}
  \country{United States}
}

\author{Thomas Synaepa-Addison}
\email{synaeptq@mail.uc.edu}
\affiliation{%
  \department{School of Information Technology}
  \institution{University of Cincinnati}
  \city{Cincinnati}
  \state{Ohio}
  \country{United States}
}

\author{Jess Kropczynski}
\email{jess.kropczynski@uc.edu}
\affiliation{%
  \department{School of Information Technology}
  \institution{University of Cincinnati}
  \city{Cincinnati}
  \state{Ohio}
  \country{United States}
}

\author{Pamela J. Wisniewski}
\email{pwisniewski@icsi.berkeley.edu}
\affiliation{%
  \department{Socio-Technical Interaction Research Lab}
  \institution{International Computer Science Institute}
  \city{Berkeley}
  \state{California}
  \country{United States}
}

\renewcommand{\shortauthors}{Ma et al.}

\begin{abstract}
Age verification is often treated as a technical problem: can a system determine a child’s age accurately? We argue this framing is too narrow. Age assurance becomes consequential when evidence is evaluated, translated into age-related claims, and used to decide whether a person can access, purchase, or belong. We review 85 publications on children’s age assurance published from 2020 through February 2026. We find that shared terms such as age verification describe different processes. Age is represented as threshold eligibility or inferred estimation, and the same eligibility claim can arise from different evidence and components. Rights, access, and privacy receive more attention than accuracy, error, and fairness; yet institutional actors are rarely connected to system failures or user remedies, an accountability gap. We introduce the \emph{Age-Assurance Process Framework}, which treats age assurance as a sociotechnical process connecting evidence, evaluation, claims, and decisions rather than reducing it to a technical problem.
\end{abstract}




\maketitle

\section{Introduction}
\label{sec:intro}

Digital services increasingly use age to determine what users may see, buy, or join. Regulators worldwide now require services to establish a user's age, age range, or eligibility threshold \cite{persson2024as,jarvie2024government,khlerdauner2025child}, assuming that accurate age determination will reduce risks to children \cite{reid2025nerd}. However, researchers question both the feasibility and the consequences of these mandates \cite{stardust2024mandatory,yar2020protecting}. Inconsistent terminology further complicates this debate. Standards and children's rights research use \emph{age assurance} as an umbrella term \cite{livingstone2024children,iso2025ageassurance}, while terms like \emph{age verification}, \emph{age estimation}, \emph{age inference}, and \emph{self-attestation} remain inconsistently defined \cite{association2024ieee}. In practice, these systems operate by requesting evidence from users. A child seeking to play a game, access a platform, or purchase a regulated product may be asked for a document, a facial image, or a date of birth. Because these requests can disclose more information than the age claim itself, requiring proof of age often means collecting identifiable data from all users \cite{scheffler2024systems,stardust2024mandatory}. Age assurance therefore compromises children's privacy, access to information, equal treatment, and eventually online safety \cite{livingstone2024children,livingstone2025there,wisniewski2024moving}. 

Prior research lacks a coherent basis for comparing how age assurance operates. System evaluation work measures how accurately models estimate age from faces or gestures \cite{anda2020assessing,hossain2023case,west2024picture,pulfrey2022zoom}. Policy audits evaluate what retailers require at checkout \cite{reile2025alcohol,egan2023absence,terala2023access,williams2020sales}. Qualitative studies explore how young people experience these age checks \cite{woodley2025australian,schiff2021accessing}, while legal analyses interpret what statutes require of digital services \cite{grimmelmann2024return,murray2025cyber,pasquale2022consent,adrogu2026playground}. Although these strands of work examine different parts of age assurance, they often use the same term, \emph{age verification}. This shared label obscures whether two studies evaluate the same input evidence, age-related claim, or downstream decision. Digital services further complicate comparison by combining multiple methods without a consistent end-to-end procedure \cite{vanderhof2022we}. Without a common unit of analysis, interdisciplinary studies remain difficult to translate into actionable guidance for design and policy.

This gap is increasingly consequential as governments translate age-assurance terminology into enforceable requirements. Australia mandates a minimum age of 16 for social media \cite{au2024smma}; the UK requires effective age assurance for pornographic content \cite{ofcom2025hea}; Texas requires app stores to verify age and obtain parental consent \cite{tx2025sb2420}; and California will require operating-system providers to provide age-bracket signals by 2027 \cite{ca2025ab1043}. These policies regulate different parts of a \emph{sociotechnical} process---turning evidence into age-related claims and decisions---while assuming age assurance can be consistently specified and audited. Yet, the literature lacks a framework to connect (1) the evidence a system accepts, (2) the evaluation method, (3) the resulting age-related claim, and (4) the downstream decision. To address this gap, our review unites system evaluations, compliance audits, qualitative studies, and legal analyses into a single corpus, analyzing the reported age assurance across these four roles. We ask three research questions:

\begin{itemize}[nosep]
    \item[\textbf{RQ1:}] \textit{How is digital age assurance for children studied across research contexts, methods, and domains?}

    \item[\textbf{RQ2:}] \textit{How does prior research conceptualize and operationalize digital age assurance for children?}

    \item[\textbf{RQ3:}] \textit{What sociotechnical challenges and gaps does prior research identify in age assurance?}
\end{itemize}

To answer these questions, we conducted a systematic literature review (SLR) of 85 publications (2020--February 2026) addressing children's age assurance across five academic databases and Google Scholar, following PRISMA guidelines \cite{page2021prismaStatement,page2021prismaExplanation}. We combined a publication-level analysis with a detailed extraction of 251 age-assurance processes. To ensure consistency across the corpus, we used LLM-assisted extraction followed by independent human validation.

We find that age-assurance research is recent, with 53 of the 85 publications appearing in 2024 or 2025, and concentrated in public health and computing. Research methods vary by domain: regulated-product studies rely on compliance audits, whereas biometric studies evaluate algorithmic systems (RQ1). Across these settings, the term \emph{age verification} refers interchangeably to document validation, algorithmic estimation, self-reported age, or a simple access rule. Age is primarily represented as threshold eligibility (N=52) or estimation from observable data (N=37). However, these representations obscure how the underlying claim was produced; a government ID and a self-reported birth date, for example, can yield the same threshold-eligibility claim. Therefore, an age claim cannot be understood apart from its evidence and its downstream decision (RQ2). Regarding sociotechnical challenges, prior work prioritizes rights and access (N=52) and privacy and data protection (N=46) over accuracy, error, and fairness (N=25). While prior work assigns operational roles to platforms, providers, regulators, and users, it rarely identifies who is responsible when a child is unjustly excluded or what recourse exists. We identify this systemic disconnect as an accountability gap (RQ3).

Our SLR makes three contributions to HCI and children's online privacy and safety: \textbf{(1) Mapping the interdisciplinary literature on age assurance:} We synthesize 85 publications across public health, computing, governance, and identity standards, demonstrating how shared terminology obscures fundamental differences in evidence, claims, and decisions.
\textbf{(2) Exposing sociotechnical challenges and accountability gaps:} We show that while prior work emphasizes rights, access, and privacy, it rarely connects system failures to responsible institutional actors or to remedies for excluded users.
\textbf{(3) Reconceptualizing age assurance as a sociotechnical process:} We introduce the \emph{Age-Assurance Process Framework} to connect input evidence, evaluation, age-related claims, and downstream decisions. This framework shifts the focus from benchmarking isolated components to evaluating whether the complete process is proportionate to the privacy it demands and the decisions it informs.

\section{Background}
\label{sec:background}

\subsection{Age Assurance as an Umbrella for Distinct Operations}
\label{sec:umbrella}

\emph{Age assurance}, \emph{age verification}, \emph{age estimation}, \emph{age inference}, and \emph{self-declaration} are related but non-equivalent terms for processes that rely on different forms of evidence, make different claims about age, and provide different levels of certainty. We use \emph{age assurance} as the umbrella term based on prior child rights and standards work \cite{livingstone2024children, iso2025ageassurance}; the terms beneath it describe different ways of establishing or asserting age:

\begin{itemize}[leftmargin=1.2em,itemsep=1pt,topsep=1pt,parsep=0pt]
    \item \textbf{Age assurance} \textit{refers to the broader set of processes used to establish or assess a person's age or age-related eligibility} \cite{livingstone2024children,iso2025ageassurance}.
    
    \item \textbf{Age verification} \textit{seeks a high-certainty determination based on documentary or other authoritative evidence, such as a government-issued identity document} \cite{association2024ieee}.
    
    \item \textbf{Age estimation} \textit{infers age from observable signals, such as facial geometry or interaction behavior, without necessarily establishing the user's identity} \cite{anda2020assessing,hossain2023case}.
    
    \item \textbf{Age inference} \textit{derives age-related information from signals a service already holds, such as the age of an account or other account metadata} \cite{iso2025ageassurance}.
    
    \item \textbf{Self-declaration} \textit{accepts the age a user states without requiring external evidence} \cite{vanderhof2022we}.
\end{itemize}

Recent standards formalize these distinctions differently. The Institute of Electrical and Electronics Engineers (IEEE) 2089.1 standard places age-verification methods on a common assurance scale beginning with an asserted age \cite{association2024ieee}. Conversely, the International Organization for Standardization and the International Electrotechnical Commission (ISO/IEC) 27566-1 standard frames age assurance around an eligibility decision, separating the policymaker who defines the requirement, the service provider who produces the result, and the relying party who acts on it \cite{iso2025ageassurance}. In this SLR, we reserve \emph{age verification} for this narrower operation or when reporting a study's original terminology. We use \emph{self-attestation} for what these sources call self-declaration, and we retain \emph{age gating} when reporting prior work that describes an access rule without specifying how the underlying age claim was established.

Outside these standards, however, the literature uses \emph{age verification} interchangeably to describe estimation from observable signals \cite{anda2020assessing,hossain2023case}, the acceptance of documentary evidence \cite{murray2025cyber,williams2020sales}, or enforcement at an interface \cite{egan2023absence,reile2025alcohol}. Meanwhile, legal research evaluates these overlapping processes against privacy \cite{scheffler2024systems}, proportionality \cite{stardust2024mandatory}, and institutional responsibility \cite{livingstone2024children}. Because a mechanism label alone does not reveal which part of an age-assurance process a publication actually studies, we treat terminology and source-specific definitions as objects of analysis.

\subsection{Age Assurance as a Sociotechnical Process}
\label{sec:sociotechnical}
An age-related decision is rarely the output of a single \emph{mechanism} (e.g., a document check or a facial estimator). For instance, when a service admits a user only after a document check, the user supplies a document, a verification provider validates it, and the service decides whether to grant access based on a received age claim. Here, four parts of one decision sit with different parties, and none holds all four. Existing standards each describe only part of this sequence. The National Institute of Standards and Technology (NIST) identity-proofing guidance treats proofing as multi-step and partly human, defining \emph{resolution}, \emph{validation}, and \emph{verification} stages to collect and confirm the authenticity and ownership of identity evidence \cite{temoshok2025nist}. However, this guidance ends at enrollment and says little about the access decisions a relying service later makes. Similarly, ISO/IEC 27566-1 separates the provider from the relying party but assigns those parts to actors rather than tracing a single decision flow \cite{iso2025ageassurance}. Meanwhile, IEEE 2089.1 ranks whole methods on one assurance scale without separating the sequence into parts \cite{association2024ieee}. Synthesizing these accounts reveals four roles: the \textbf{\emph{input evidence or signal}} a user supplies or a system observes, the \textbf{\emph{evaluating component}} that acts on it, the \textbf{\emph{age-related claim}} it establishes, and the \textbf{\emph{downstream decision}} that claim informs. We use the term \emph{age-assurance process} for one complete sequence across these four roles, and it is the primary analytic unit our SLR records. Whether reviewed publications specify or omit these specific roles remains an empirical question addressed in Section~\ref{sec:rq2-processes}.

Three theoretical traditions justify placing the boundary of our analysis beyond the technical mechanism. First, sociotechnical systems theory holds that a technical mechanism cannot be evaluated apart from the social and institutional context that gives it its effect \cite{baxter2011socio}. Second, the human-in-the-loop framework applies that principle to security, treating human communication, attention, and action as core components of a secure system rather than as external sources of error \cite{cranor2008human}. Finally, the security ceremony concept extends the protocol boundary past the machines to the interfaces, workflows, and physical transfers of data-bearing objects around them \cite{ellison2007ceremony}. Because an age-assurance process similarly distributes work across users, operators, verification providers, and regulators, the interfaces where these parties intersect are integral parts of the system.

This expanded boundary matters most when the decision falls on a child. Safety-driven access restrictions directly impact children's expression, participation, privacy, and access \cite{uncrc2021digital,livingstone2024children}. Because parents and guardians often authorize access or support account recovery \cite{vanderhof2022we,pasquale2022consent}, these processes must establish both an adult--child relationship and an age. Furthermore, when a process wrongly refuses a young user, the explanation and any route to contest the refusal depend on the process as a whole, not just on the algorithmic component that made the error. The boundary also determines how the literature should be reviewed. Because two publications using the same mechanism label may study entirely different processes, our SLR organizes its synthesis by recording which evidence, component, claim, and decision each publication connects, rather than relying on inconsistent mechanism labels.

\section{Methods}
\label{sec:methods}

\subsection{A Systematic Literature Review Using PRISMA}
Our SLR combines two levels of analysis. The publication level characterizes what each publication studies and how; the process level records each age-assurance process a publication reports as a \emph{four-part} record, one part for each of the four roles derived in Section~\ref{sec:sociotechnical}. We combined them because the literature's methods do not produce equivalent evidence: the publication level records those differences, and the process level supplies a unit comparable across them. We report identification and study selection with the PRISMA diagram and reporting items \cite{page2021prismaStatement,page2021prismaExplanation}; Appendix~\ref{app:search} records where our procedure departs from them. 

\begin{figure}
  \centering
  \includegraphics[width=0.65\textwidth]{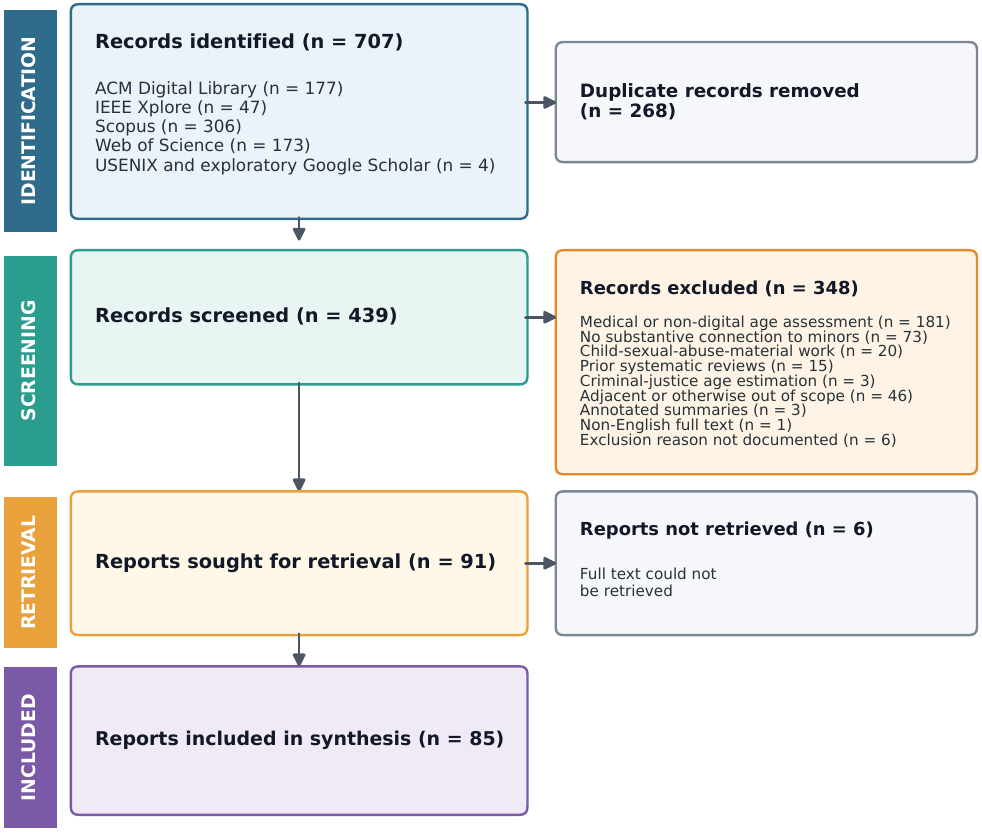}
  \caption{PRISMA flow diagram from 707 identified records to 85 included publications.}
  \Description{A four-stage flow diagram shows 707 records identified from six search sources, with USENIX and exploratory Google Scholar results combined in one reporting line. After removing 268 duplicates, we screened 439 unique records. The exclusion box lists 181 medical or non-digital age-assessment records, 73 records without a substantive connection to minors, 20 child-sexual-abuse-material records, 15 prior systematic reviews, three criminal-justice age-estimation records, 46 adjacent or out-of-scope records without a formal criterion code, three annotated summaries, one non-English full text, and six records without a documented exclusion reason. These reasons sum to 348 exclusions. We sought 91 reports for retrieval; six were not retrieved, and 85 were included in the synthesis.}
  \label{fig:prisma-flow}
\end{figure}

\subsection{Search Strategy and Study Selection}
\label{sec:search}
We developed the search strategy iteratively in February 2026. We piloted terms in the ACM Digital Library, IEEE Xplore, and USENIX, then extended coverage to Scopus and Web of Science, and used Google Scholar to find publications not indexed in those databases. The common query joined three concept blocks with Boolean AND: an age-assurance block, a minor-related block, and a digital-context block (Appendix~\ref{app:search}). We applied five inclusion criteria. A publication had to:

\begin{itemize}[leftmargin=1.2em,itemsep=1pt,topsep=1pt,parsep=0pt]
  \item be fully accessible for the research team to retrieve;
  \item be in English, given the research team's language proficiency;
  \item be published in or after 2020, when the UK Age Appropriate Design Code \cite{ico2020aadc} and later child-safety mandates in the US \cite{tx2025sb2420,ca2025ab1043} and Australia \cite{au2024smma} prompted a new wave of age-assurance research (the search closed in February 2026);
  \item evaluate, design, or analyze digital age verification, estimation, or gating rather than mention age assurance in passing, a criterion that technical evaluations and legal/policy analyses could both meet; and
  \item address an age-related decision that explicitly distinguishes or protects individuals below an age of majority (e.g., under-13, under-16, or under-18 thresholds). This criterion excludes general adult identity verification, such as financial Know Your Customer, that does not address child safety, privacy, or parental mediation.
\end{itemize}

Figure~\ref{fig:prisma-flow} summarizes study selection. Searches across the five databases and Google Scholar yielded 707 records, and relevance screening of the 439 unique records left 91 candidates for full-text retrieval. The largest exclusion categories were clinical, forensic, or non-digital age assessments such as bone or dental X-rays (n=181), absence of a substantive focus on minors or children's digital access (n=73), and adjacent or out-of-scope digital governance topics and other final-eligibility exclusions (n=56). The screening ledger in Appendix~\ref{app:search} reports every category, including prior systematic reviews (n=15). Six of the 91 could not be retrieved, giving a final corpus of 85 publications. Relevance screening examined available full texts rather than abstracts alone. One researcher performed the primary screening; a second researcher independently reviewed metadata, verified ambiguous records, and refined search queries; and the research team resolved borderline cases by consensus.

\subsection{Extracting and Structuring Age-Assurance Processes}
\label{sec:extraction}

The extraction protocol defined 17 evidence fields and one optional synthesis memo (Table~\ref{tab:field-families}). The fields cover the research setting, terminology, methods, actors, and reported concerns at the publication level and the four roles of each reported process. Every extracted value was linked to a source excerpt, its page or section location, and a rationale for why the excerpt supported it. Process-level extraction used the four roles derived in Section~\ref{sec:sociotechnical}, defined in advance:

\begin{itemize}[leftmargin=1.2em,itemsep=1pt,topsep=1pt,parsep=0pt]
    \item \textbf{Input evidence or signal:} what a user provides or a system observes.
    \item \textbf{Evaluating component:} the technical, human, or institutional component that evaluates or applies that input.
    \item \textbf{Age-related claim:} what the evaluation establishes about a person's age or eligibility.
    \item \textbf{Downstream decision:} the action or policy decision informed by that claim.
\end{itemize}

These roles offer a common description to processes that prior work labels differently. Documentary or self-reported evidence can establish eligibility against an age threshold and inform an access decision \cite{association2024ieee,reile2025alcohol}, while biometric or behavioral signals can be evaluated to estimate age \cite{hossain2023case,west2024picture}. Prior work reports both as ``age verification''; the four-part record distinguishes them by evidence, claim, and decision.

We recorded a process only when a publication connected its elements within the same mechanism, method, system description, requirement, or closely related explanation. A process did not need all four roles. If a source described an age-related claim but did not state what happened next, the downstream decision remained unspecified, so missing roles are part of the evidence rather than filled by the extraction schema. We restricted the \emph{age-related claim} role to outputs that establish a numeric age, age band, threshold-eligibility status, or formal age-verification status; other outputs, including parental consent, relationship verification, and identity matching, were recorded separately.

Because the protocol required extracting the same detailed fields across the full corpus, we used an automated pipeline built on OpenAI's GPT-5.4, followed by independent human validation (Section~\ref{sec:validation}). The pipeline accepted machine-readable text only and could record only evidence present in that text. Appendix~\ref{app:fields} reports the model parameters, the chunking and merge procedures for long publications, the retained outputs, and the extraction prompt. We disclose the model's role in data extraction in accordance with ACM policy \cite{acm2025aiPolicy}.

\subsection{Interpreting and Coding the Extracted Evidence}
\label{sec:coding}

The extracted evidence did not determine the categories reported in our findings. We interpreted recurring patterns across the corpus from the source excerpts, publication context, and bibliographic metadata, and developed categories that support comparison across research traditions.

For RQ1, we recorded publication year and first-author institutional affiliation from the full texts. We assigned each publication to one of six \emph{study-context} categories, the research field in which it appears, such as public health, computing, governance, or identity standards (Figure~\ref{fig:rq1-bibliographic}B). We followed a documented priority scheme based on titles, venues, and publication types (Appendix~\ref{app:codebook}). Other publication-level characteristics, including topical domains (what a publication examines) and modes of inquiry (how it substantiates its findings), were non-exclusive, because publications often addressed more than one topic or analytical approach. Repeated instances of the same category within a publication were counted once.

For RQ2, we first coded the extracted age-related claims into four types: threshold eligibility, numeric-age estimates, age-band estimates, and age-verification status. From these types we identified two broader representations of age: eligibility relative to a threshold, and an inferred estimate. We then interpreted each claim in relation to the evidence used to produce it, the component that evaluated that evidence, and the decision the claim informed, to compare processes reported under the same term. We also distinguished the publication-level \emph{decision purpose} for which age information is produced or used from the process-level \emph{downstream decision} that follows a particular age-related claim. A broad purpose such as digital access or compliance can encompass several operational decisions. Four processes that fit none of the resulting downstream-decision categories were retained as residual cases.

For RQ3, we distinguished \emph{challenges}, the concerns a publication discusses within its broader analysis, from \emph{research gaps}, the limitations or open questions its authors explicitly identified as remaining research needs. We coded actors from the dedicated actor field and from descriptions of social, human, or institutional components. We then compared the problems publications identify with the actors they describe as operating, governing, or participating in age-assurance processes, and whether those actors were connected to failures or remedies.

Category construction was iterative: a codebook of 111 documented coding rules translates recurring language in the publications into comparable analytical categories (Appendix~\ref{app:codebook}, which also reports how generic, residual, and unspecified values were handled). The rules formalize interpretive decisions made during category development so that they apply consistently across the corpus. An AI coding assistant (Claude) helped draft and refine the rules; the research team made all interpretive decisions. When source wording was ambiguous or fit no existing category, the lead author reviewed the supporting excerpt and resolved the case against the documented category definitions and priorities. 


\subsection{Synthesis and Validation}
\label{sec:synthesis}

\subsubsection{Cross-Corpus Synthesis between Quantitative and Qualitative Approaches}

We synthesized patterns across the corpus using descriptive counts, cross-tabulations, and close reading of the supporting excerpts. For RQ3, we also calculated exploratory phi ($\phi$) coefficients across the 120 pairs of 10 challenge categories and 12 publication-level component indicators to examine whether particular challenges co-occur with particular components (Equation~\ref{eq:phi} in Appendix~\ref{app:analysis}). Fifty-eight positive associations met our support threshold ($n \ge 5$ co-occurrences; marginal totals $\ge 8$); we excluded one association inflated by overlapping familial terminology and report the 12 largest remaining coefficients in Figure~\ref{fig:rq3-challenges-actors}B. These coefficients are descriptive associations, not evidence of causal relationships. Returning to the texts, we asked why categories co-occurred, how the same terminology was used across settings, and where similar age-assurance processes carried different meanings or consequences. Publications cited in the findings illustrate these patterns and are not a separate qualitative subsample. Because the corpus spans algorithmic benchmarks, compliance audits, qualitative studies, and statutory analysis, no single quality-appraisal instrument applies across all publications, and we did not weight publications by study design or methodological quality. Counts therefore indicate the prevalence of research attention, not the strength of evidence for a particular claim.

\subsubsection{Human Validation of LLM-Assisted Extraction}
\label{sec:validation}
We validated the LLM-assisted extraction before using it for coding or synthesis. Automated integrity checks confirmed that all 85 publication identifiers were unique, that every populated field was linked to a verbatim source excerpt, and that summary counts could be reproduced. The checks also flagged three multi-item entries supported by overly broad excerpts and ten minor quotation discrepancies for human review. Two researchers then validated the complete extraction dataset of 1,445 entries across 17 evidence fields and all 85 publications. Each independently audited half of the corpus against the original full texts, checking whether the extracted value was supported, whether relevant evidence had been missed, and whether the paired excerpt represented the source. The research team adjudicated the resulting discrepancies. 
Against the adjudicated human record, 96.7\% of extracted entries (1,398/1,445; 95\% Clopper--Pearson CI [95.7, 97.6]) required no correction, and no entry was judged unverifiable. The 47 corrections included 24 missed evidence instances, 17 value or excerpt refinements, and 6 unsupported extractions. Model--human agreement on whether evidence was present reached $\kappa = 0.81$ (95\% CI [0.75, 0.88]). The four-part process records central to RQ2 were retained without correction across 83 of the 85 publications (97.6\%). The research team incorporated all corrections before producing the reported findings, tables, and figures. Appendix~\ref{app:analysis} reports validation outcomes by field, the derivation of $\kappa$, and precision and recall.

\section{Findings}
\label{sec:findings}

\begin{figure*}[t]
  \centering
  \includegraphics[width=0.5\textwidth]{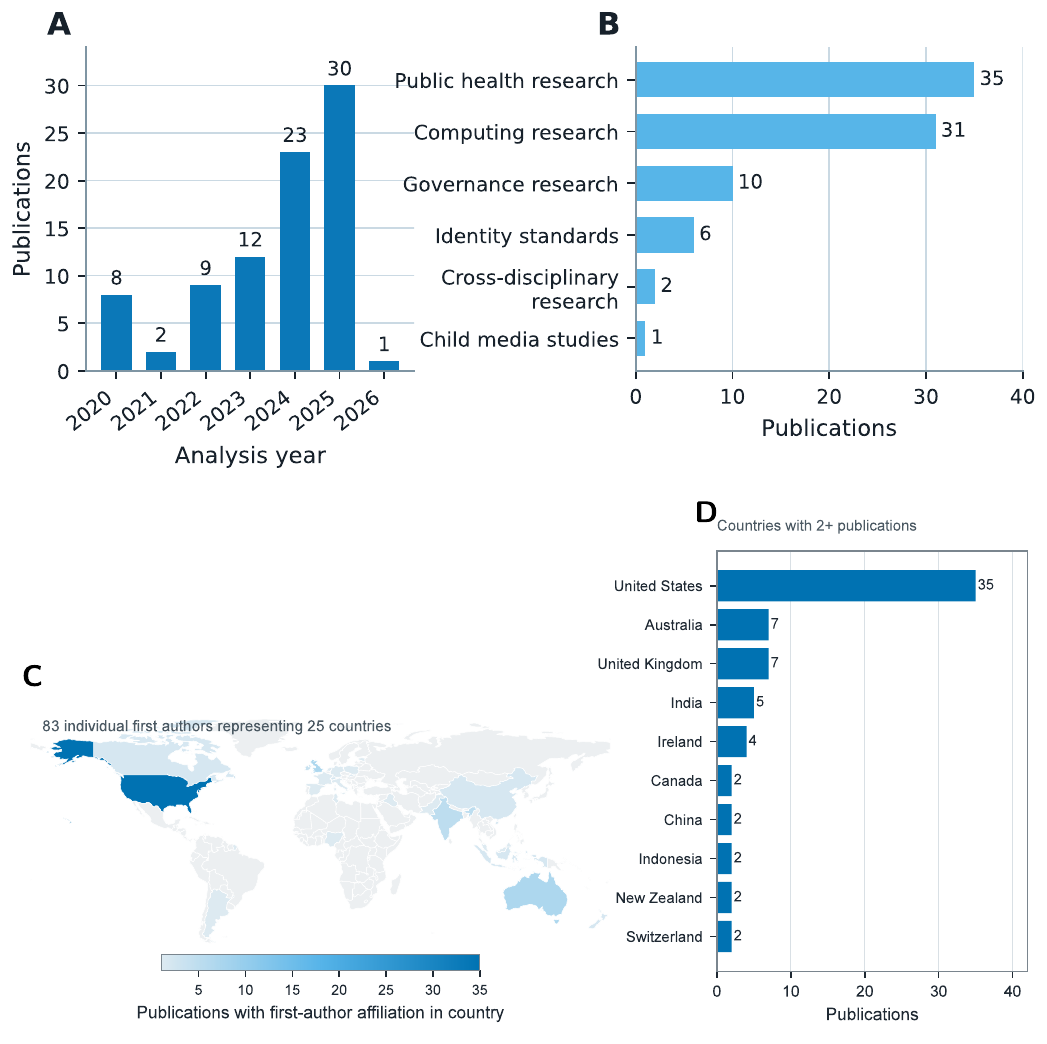}
  \caption{Publication timing, context, and first-author geography. Panels A--B report analysis year (the publication year recorded during screening) and study context; 2026 is a partial year. Panels C--D report the 25 countries among 83 individually authored publications with a reported first-author affiliation; Panel D shows only those with at least two publications.}
  \Description{Four panels. Panel A is a histogram of analysis year, rising to 23 publications in 2024 and 30 in 2025, with one partial-year record in 2026. Panel B is a bar chart of six study-context groups: public health research 35, computing research 31, governance research 10, identity standards research 6, cross-disciplinary research 2, and child media studies 1. Panel C is a world map shading each country by its number of first-author affiliations. Panel D is a bar chart of the countries with at least two publications, led by the United States with 35, then Australia and the United Kingdom with 7 each, India with 5, and Ireland with 4.}
  \label{fig:rq1-bibliographic}
\end{figure*}

\subsection{RQ1: Recent Literature Spans Research Domains and Methods}
\label{sec:rq1}

\subsubsection{Publication Years, Study Contexts, and Geography}
\findingclaim{The reviewed literature is recent, with most publications appearing in the two most recent complete years of our search window.} Publications from 2024 and 2025 account for 27\% (N=23) and 35\% (N=30) of the corpus, or 62\% (N=53) together (Figure~\ref{fig:rq1-bibliographic}A). Publication years reflect screening records, not publisher online-first metadata, which differ for 16\% (N=14) of publications (Appendix~\ref{app:codebook}).

\findingclaim{Study context places the large majority of publications in public health or computing research.} The six mutually exclusive \emph{study-context} categories describe where research appears, not authors' disciplinary identity. Public health research at 41\% (N=35) and computing research at 36\% (N=31) together hold 78\% (N=66) of publications (Figure~\ref{fig:rq1-bibliographic}B). The remaining publications appear in governance research at 12\% (N=10), identity standards research at 7\% (N=6), cross-disciplinary research at 2\% (N=2), and child media studies at 1\% (N=1). Public health research, published mainly in tobacco, alcohol, and adolescent-health journals, most often audits online sellers by attempting purchases or coding their websites. These audits record whether a checkout asks for more than a tick-box or a date of birth \cite{egan2023absence,terala2023access}, and whether a courier requests identification at delivery \cite{reile2025alcohol,sneyd2024alcohol}. Computing research, by contrast, most often treats age as a quantity to estimate from a face \cite{anda2020assessing,bao2023general}, a voice \cite{durgam2024estimation}, or a touch gesture \cite{hossain2023case,pulfrey2022zoom} and reports accuracy or robustness to replay attacks \cite{korshunov2024vulnerability}.

\findingclaim{First-author affiliations span many countries, yet the US accounts for roughly two-fifths of publications.} We identified first-author countries for 83 of the 84 individually authored publications; one group-authored standards publication was excluded, and one book chapter omitted affiliation details. Across these 83 publications, affiliations cover 25 countries, led by the US at 42\% (N=35), then Australia and the UK at 8\% (N=7) each (Figure~\ref{fig:rq1-bibliographic}C--D). This distribution reflects institutional affiliation rather than study population, author nationality, or global deployment prevalence, and is bounded by our English-language inclusion criterion.

\subsubsection{Research Objects and Evidence}
\begin{figure}[tb]
  \centering
  \includegraphics[width=0.5\textwidth]{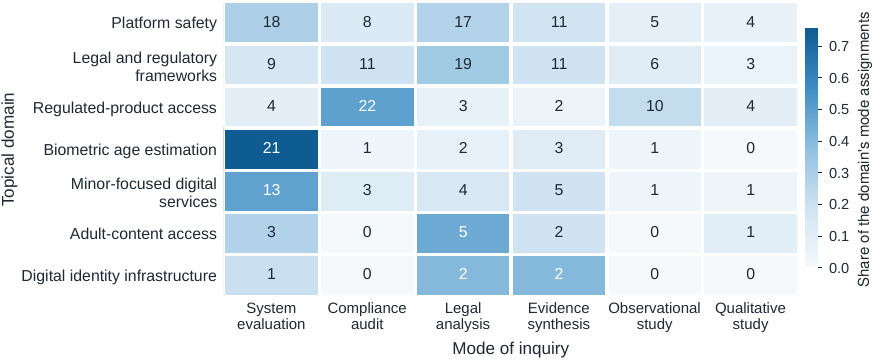}
  \caption{Topical and inquiry composition of the 85 publications. Each cell counts unique publications and is shaded by that count's share of the domain's mode-of-inquiry assignments. Domains and modes are non-exclusive, so the counts describe topic coverage rather than exclusive groups.}
  \Description{A heatmap crossing seven topical domains with six modes of inquiry. Each cell prints a publication count and is shaded by that count's share of the domain's mode-of-inquiry assignments. Platform safety is highest in system evaluation (18) and legal analysis (17); legal and regulatory frameworks is highest in legal analysis (19); regulated-product access is highest in compliance audit (22); biometric age estimation is highest in system evaluation (21); and minor-focused digital services is highest in system evaluation (13).}
  \label{fig:rq1-architecture}
\end{figure}

Study contexts describe disciplinary venues, whereas seven non-exclusive \emph{topical domains} describe what publications examine within age assurance (Figure~\ref{fig:rq1-architecture}).
\findingclaim{The two most prevalent topical domains are platform safety and legal and regulatory frameworks, whereas adult-content access and digital identity infrastructure are the least common:}
(1)~\emph{platform safety}, 52\% (N=44), examining age assurance on social media, gaming, and other online platforms;
(2)~\emph{legal and regulatory frameworks}, 44\% (N=37), analyzing age-assurance mandates and constitutional implications;
(3)~\emph{regulated-product access}, 35\% (N=30), auditing age assurance for online sales of alcohol, tobacco, cannabis, and e-cigarettes;
(4)~\emph{biometric age estimation}, 28\% (N=24), evaluating facial, vocal, or behavioral estimators;
(5)~\emph{minor-focused digital services}, 24\% (N=20), examining online platforms designed primarily for minors;
(6)~\emph{adult-content access}, 7\% (N=6), analyzing age assurance for restricted adult media; and
(7)~\emph{digital identity infrastructure}, 4\% (N=3), developing privacy-preserving credentials.
Platform-safety research examines how social media apps establish a user's age, from a date of birth entered at sign-up to the selfie or identity document that verifies it \cite{eltaher2025loophole,vanderhof2022we}. Legal and regulatory research analyzes mandates under free-speech law \cite{grimmelmann2024return} and maps state laws on e-cigarette delivery sales \cite{azagba2023loopholes} and child online safety \cite{reid2025nerd}.

\findingclaim{The corpus draws on six non-exclusive modes of inquiry, and which one dominates shifts with the topical domain.} \emph{Modes of inquiry} describe the methods that prior work adopts to substantiate its findings: system evaluation, 42\% (N=36); compliance audit, 31\% (N=26); legal analysis, 27\% (N=23); evidence synthesis, 18\% (N=15), which assembles existing law, policy, or standards as a primary contribution; observational study, 14\% (N=12); and qualitative study, 8\% (N=7). Across domains (Figure~\ref{fig:rq1-architecture}), work on regulated-product access relies on compliance audits and shopper observation: one mystery-shopping study placed test alcohol orders in Estonia and recorded whether identification was requested at delivery \cite{reile2025alcohol}. Work on biometric age estimation instead uses system evaluations that measure commercial cloud estimators against a dataset of underage faces with known ages \cite{anda2020assessing}. Research on adult-content access centers on statutory interpretation rather than empirical testing \cite{grimmelmann2024return,yar2020protecting}.

\findingclaim{The reviewed publications rarely present an age-assurance process as purely technical.} These component categories record what a publication discusses anywhere in its text, so they are broader than the single \emph{evaluating component} that occupies one role in a process (Section~\ref{sec:rq2-processes}). \emph{Technical components} are algorithmic, hardware, or data-processing operations that estimate age, evaluate evidence, or apply an age-related rule; \emph{social, human, or institutional components} are actions people or organizations perform within an age-assurance process. Both kinds appear in 91\% (N=77) of publications, often within one process. An online tobacco seller, for example, may pass a date of birth or driver's license image to a verification service and rely on a delivery driver to check the recipient's identification \cite{williams2020sales}. A social app may pair a date-of-birth gate with a parental-consent flow \cite{pasquale2022consent,vanderhof2022we}, and a service that estimates age from a selfie may still require identification from the users its estimator flags \cite{scheffler2024systems}. Technical components alone appear in 8\% (N=7) of publications, and institutional components alone in 1\% (N=1).

\findingclaim{An explicit legal or policy context appears in most publications, with regulated-product sales being the most frequent.} A legal or policy context is explicit in 89\% (N=76) of publications. Of those 76, 82\% (N=62) fall under one of five named policies and 18\% (N=14) match none. Across the 85 publications, the three most frequent of these overlapping settings are regulated-product sales at 35\% (N=30), child privacy and consent frameworks at 29\% (N=25), and platform and online safety legislation at 22\% (N=19). These differences in study context, domain, mode, and policy context shape what prior work means by ``age verification'' (Section~\ref{sec:rq2}).

\begin{tcolorbox}[enhanced, colback=cognitiveblue!6, colframe=cognitiveblue!6, boxrule=0pt, arc=4pt, fuzzy shadow={2pt}{-2pt}{0pt}{0.2pt}{black!15}]
\textbf{\textcolor{cognitiveblue}{Key Takeaway:}} 
The literature is concentrated in public health and computing, relying on distinct methods like compliance audits and system evaluations. Because these domains use a shared term like ``age verification'' to describe different operations, a mechanism label alone obscures what a publication actually studies.
\end{tcolorbox}

\subsection{RQ2: How Age Assurance Is Conceptualized and Operationalized}
\label{sec:rq2}.

\subsubsection{The Same Terms Refer to Different Operations}
\label{sec:rq2-term}

\begin{figure}[tb]
  \centering
  \includegraphics[width=0.5\textwidth]{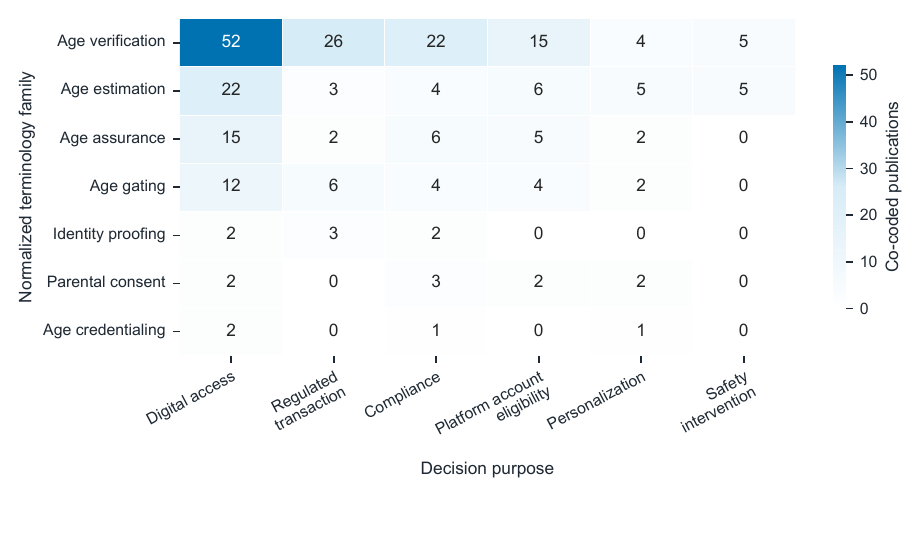}
  \caption{Publication-level overlap between terminology and decision purpose. Each cell counts unique publications. Terminology families and decision purposes are both non-exclusive, so a row can sum to more than the publications that use the term.}
  \Description{A heatmap crossing seven terminology families with six decision purposes, each cell counting unique publications. Age verification is the largest row, reaching 52 publications under digital access, 26 under regulated transaction, 22 under compliance, and 15 under platform-account eligibility.}
  \label{fig:rq2-terminology-definitions}
\end{figure}

\findingclaim{While ``age verification'' is the most common term, appearing in roughly three-quarters of publications, it refers to several distinct operations.}
\emph{Age verification} appears in 74\% (N=63) of publications, \emph{age estimation} or prediction in 29\% (N=25), \emph{age assurance} in 19\% (N=16), and \emph{age gating} in 14\% (N=12), and these terms overlap within publications. What each term denotes depends on the part of the process a publication examines. Publications on age estimation infer age from observable signals such as touch gestures \cite{hossain2023case} or facial imagery \cite{anda2020assessing}. Legal and standards publications use verification or assurance to discuss documentary evidence, eligibility, privacy, and institutional governance \cite{livingstone2024children,association2024ieee}. Publications using \emph{age gating} often describe an access rule without specifying how the underlying age claim was established \cite{persson2024as}. The same terminology can therefore refer to the input evidence, the evaluating component, the resulting age-related claim, or the downstream decision (Figure~\ref{fig:rq2-terminology-definitions}).

\findingclaim{Prior work defines age assurance primarily through what a system does rather than through a consistent formal concept.}
Among the 73 publications that define or operationally describe age assurance, 51\% (N=37) characterize it through downstream access control, 44\% (N=32) through algorithmic estimation, 41\% (N=30) through identity or documentary validation, and 32\% (N=23) through threshold eligibility. These categories overlap because one age-assurance process can contain several operations. A publication may describe document validation as the procedure, threshold eligibility as the resulting claim, and restricted entry as the decision \cite{association2024ieee,livingstone2024children,vanderhof2022we}. Definitions centered on access control, however, describe what is enforced while often leaving unclear how age was established. Audits of online cannabis and cannabidiol sellers define the check by where it sits---a confirmation before products can be viewed, and an identification request at delivery \cite{terala2023access,egan2023absence}. Definitions centered on estimation or document validation describe how age is assessed while sometimes leaving its downstream use unspecified \cite{pasquale2022consent,vanderhof2022we}.

\subsubsection{Age Functions as Population, Boundary, and Measurement}
\label{sec:rq2-role}

\findingclaim{Age-group information serves three non-exclusive functions, while the thresholds used for decisions are largely set by policy rather than developmental stages.}
Across the corpus (Appendix Figure~\ref{fig:rq2-age-operationalization}A, left), age functions as a \emph{broad population reference} identifying who a publication concerns, 96\% (N=82); a \emph{threshold-defined age category} establishing an institutional decision boundary, 59\% (N=50); or a \emph{sample or model-evaluation grouping} used to compare performance, 29\% (N=25).

The threshold-defined function depends on a cut-point, and explicit cut-points concentrate around four ages (Appendix Figure~\ref{fig:rq2-age-operationalization}A, right): age 18 at 38\% (N=32), age 13 at 24\% (N=20), age 21 at 8\% (N=7), and age 16 at 7\% (N=6). Their meaning comes primarily from the institutional context in which they are enforced. Age 13 reflects child-privacy requirements under COPPA \cite{pasquale2022consent}, age 16 appears in emerging social-media restrictions \cite{au2024smma}, age 18 commonly marks legal adulthood \cite{stardust2024mandatory,reile2025alcohol}, and age 21 governs access to regulated products such as cannabis or alcohol \cite{rhee2025gating,azagba2023loopholes}.

\subsubsection{Age Is Represented as Eligibility or Estimation}
\label{sec:rq2-rep}

The function an age group serves is separate from the form of the age claim a system acts on. \findingclaim{Publications represent age primarily as threshold eligibility or inferred estimation.}
An \emph{eligibility} representation, 61\% (N=52), establishes whether a user falls on the permitted side of a threshold \cite{association2024ieee}. An \emph{estimation} representation, 44\% (N=37), infers a numeric age, age band, or threshold probability from observable data \cite{west2024picture}: one publication predicts a smartphone user's exact age from zoom gestures and leaves the downstream decision unspecified \cite{hossain2023case}. Of the 85 publications, 41\% (N=35) address only eligibility, 24\% (N=20) only estimation, 20\% (N=17) both, and 15\% (N=13) fall under other or unspecified representation (Appendix Figure~\ref{fig:rq2-age-operationalization}B, left).

A probabilistic estimate may determine access, but it is neither an authoritative credential nor a final admission decision; a threshold-eligibility claim, in turn, does not reveal whether it came from an identity document, self-attestation, algorithmic estimation, or human review.

Both representations serve the same dominant purpose, yet they diverge beneath it (Appendix~\ref{app:visualizations}, Figure~\ref{fig:rq2-purpose}). Digital access leads for both, at 79\% (N=41) of the 52 eligibility publications and 81\% (N=30) of the 37 estimation publications. Below that, eligibility claims support regulated transactions far more often than estimates do, 46\% (N=24) against 8\% (N=3), whereas estimates more often support safety intervention, 14\% (N=5) against 4\% (N=2).

\subsubsection{Evidence and Evaluation Produce Age-Related Claims}
\label{sec:rq2-processes}

We traced 251 age-assurance processes across the 85 publications through the four roles: (1)~the input evidence or signal, (2)~the evaluating component, (3)~the age-related claim, and (4)~the downstream decision. Of these processes, 63\% (N=159) specify all four roles, and fully specified processes appear in 86\% (N=73) of the 85 publications. When publications leave a role unstated, it remains unspecified in our analysis; the downstream decision is omitted most often.

\findingclaim{Which evaluating component is most prevalent depends on whether publications name the component or describe the check only generically.}
The generic-wording rule in Appendix~\ref{app:codebook} classifies generically described checks from the evidence or output involved, so we report two counts. Among processes that explicitly name the component, algorithmic age estimation appears most often, in 35\% (N=30) of the 85 publications, followed by platform account gates at 19\% (N=16), documentary proofing at 12\% (N=10), and self-attestation at 11\% (N=9). With generic descriptions classified as well, documentary proofing becomes the most prevalent component, appearing in 49\% (N=42), followed by algorithmic age estimation at 40\% (N=34), platform account gates at 25\% (N=21), and self-attestation at 24\% (N=20). The gap is largest for documentary proofing, which publications most often describe without naming the component.

\begin{figure}[tb]
  \centering
  \includegraphics[width=\textwidth,height=0.45\textheight,keepaspectratio]{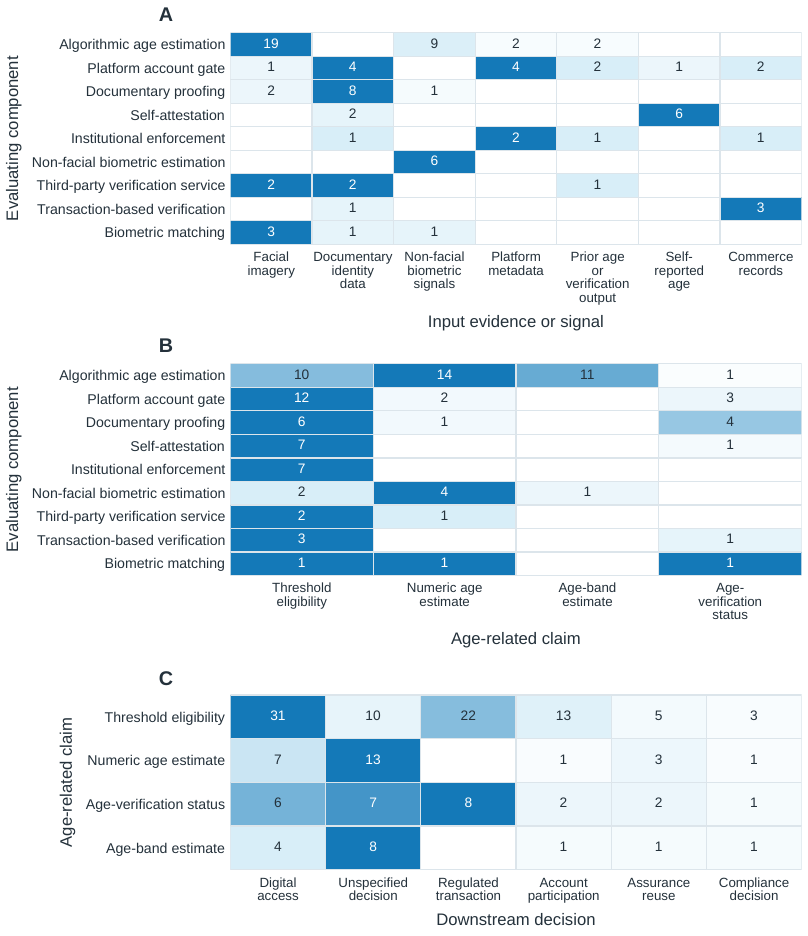}
  \caption{Frequent evaluating component--input, component--claim, and claim--decision links. Panels A and B use the 138 processes whose source text explicitly identifies the component category; Panel C draws on all 251 processes. Each panel shows the most frequently recorded categories rather than every category; residual categories are omitted, and where two categories tie at the cutoff for display, the panel shows one of them. Shading is scaled within each row, so color compares cells along a row and not between rows. Counts are unique publications and do not compare performance.}
  \Description{Three heatmaps display selected frequent coded links. The first crosses evaluating components with input evidence or signals, the second crosses components with age-related claims, and the third crosses claims with downstream decisions. Documentary proofing with documents, age estimation with facial imagery, self-attestation with self-report, and platform account gates with platform metadata are prominent cells.}
  \label{fig:rq2-linkages}
\end{figure}

These four evaluating components connect different forms of evidence to different claims (Figure~\ref{fig:rq2-linkages}A--B). \emph{Documentary proofing} evaluates documentary records, asserted identity, and issuer authority \cite{association2024ieee,crepax2022information}. For example, an adult-content service may accept a government-issued identification document or financial instrument \cite{murray2025cyber}. \emph{Algorithmic age estimation} converts facial, vocal, touch, or behavioral signals into a numeric age, age band, or threshold prediction \cite{hossain2023case}; one case study describes TikTok continuously analyzing camera frames to detect faces and predict age \cite{west2024picture}. \emph{Platform account gates} apply stored platform states or rules \cite{odeigah2025underage,eltaher2025loophole}, while \emph{self-attestation} treats a user's stated age as evidence \cite{vanderhof2022we}.

Within one process, the four roles remain distinct: a government document is what a documentary check evaluates \cite{association2024ieee,reile2025alcohol}, and facial imagery is evidence supplied to an estimator rather than the evaluating component \cite{west2024picture}.

\findingclaim{Different forms of evidence can produce the same age-related claim.}
Documentary identity data are the most frequently described input, appearing in 55\% (N=47) of publications, followed by facial imagery at 40\% (N=34), self-reported age at 22\% (N=19), and non-facial biometric signals at 18\% (N=15). Other processes use platform metadata, commerce records, prior age or verification outputs, cryptographic credentials, parent- or guardian-related evidence, or human observation.

\begin{figure*}[tb]
  \centering
  \includegraphics[width=0.6\textwidth]{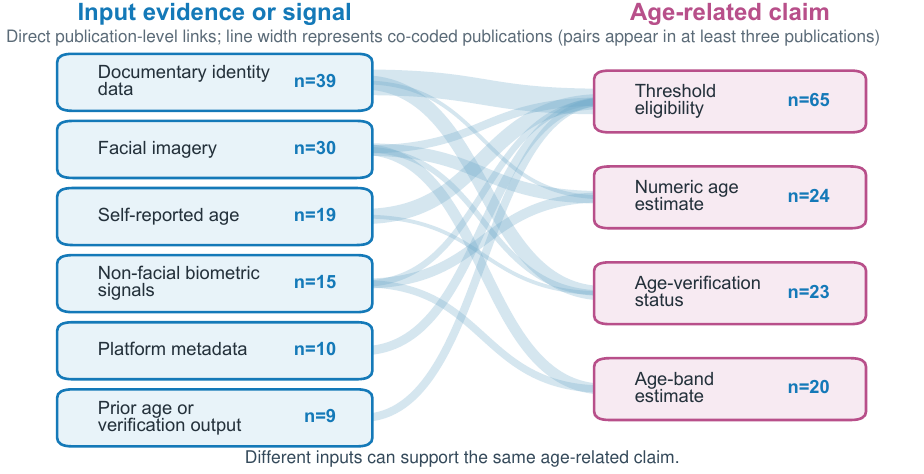}
  \caption{Direct links between input evidence and four age-related claims. The view uses the 205 processes whose recorded output is one of these claims, drawn from 83 of the 85 publications; authorization, relationship, consistency, corroboration, and residual outputs are excluded. Node counts are unique publications within that subset. The view shows the six most frequently recorded input categories, so residual and low-frequency inputs are omitted, and bands appear for pairs occurring in at least three publications.}
  \Description{A Sankey-style diagram connects documentary identity data, facial imagery, self-reported age, non-facial biometrics, platform metadata, and prior age or verification outputs to threshold eligibility, numeric-age estimates, age-verification status, and age-band estimates.}
  \label{fig:rq2-signal-claim}
\end{figure*}

Documentary identity data, self-reported age, and facial imagery all support threshold eligibility through different processes (Figure~\ref{fig:rq2-signal-claim}). Across the 85 publications, documentary identity data most often support threshold eligibility, 33\% (N=28), or age-verification status, 18\% (N=15). Facial imagery supports numeric-age estimates, 15\% (N=13), age-band estimates, 13\% (N=11), and threshold eligibility, 9\% (N=8). Self-reported age most often supports threshold eligibility, 19\% (N=16), while non-facial biometric signals primarily support numeric-age estimates, 12\% (N=10), and age-band estimates, 8\% (N=7). The label ``age verification'' can therefore describe a status grounded in institutional evidence \cite{association2024ieee}, a probabilistic estimate \cite{hossain2023case}, or a self-report accepted by an access rule \cite{vanderhof2022we}.

\subsubsection{Claims Take Meaning from Their Decision Context}
\label{sec:rq2-decisions}

Threshold eligibility is the most frequent age-related claim in the process-level analysis, appearing in 76\% (N=65) of publications, followed by numeric-age estimates at 28\% (N=24), age-verification status at 27\% (N=23), and age-band estimates at 24\% (N=20). These counts differ from the publication-level representation counts above. The claim counts are non-exclusive and record every claim type a publication's processes produce.

\findingclaim{An age-related claim cannot be interpreted apart from the evidence that produced it and the decision it informs.}
Threshold eligibility most often informs digital access, followed by regulated transactions and account participation \cite{odeigah2025underage,reile2025alcohol}. One such claim can also rest on different evidence: in the alcohol mystery-shopping study, threshold eligibility rested on a tick-box at store entry and, at delivery, on the courier's identification check \cite{reile2025alcohol}. For numeric-age and age-band estimates, prior work frequently leaves the downstream decision unspecified; when one is stated, it is most often digital access (Figure~\ref{fig:rq2-linkages}C) \cite{hossain2023case}. Across these findings, age assurance is not operationalized through a single mechanism or representation of age.

\begin{tcolorbox}[enhanced, colback=cognitiveblue!6, colframe=cognitiveblue!6, boxrule=0pt, arc=4pt, fuzzy shadow={2pt}{-2pt}{0pt}{0.2pt}{black!15}]
\textbf{\textcolor{cognitiveblue}{Key Takeaway:}} 
Age is primarily represented as threshold eligibility or inferred estimation, but the exact same age-related claim can arise from vastly different evidence. Therefore, an age claim cannot be evaluated in isolation; it must be understood as a complete sociotechnical process connecting input evidence, the evaluating component, the claim, and the downstream decision.
\end{tcolorbox}

\subsection{RQ3: Challenges Extend Beyond Accuracy Across Sociotechnical Processes of Age Assurance}
\label{sec:rq3}


\subsubsection{Challenges Across Processes}

\begin{figure*}[t]
  \centering
  \includegraphics[width=0.45\textwidth]{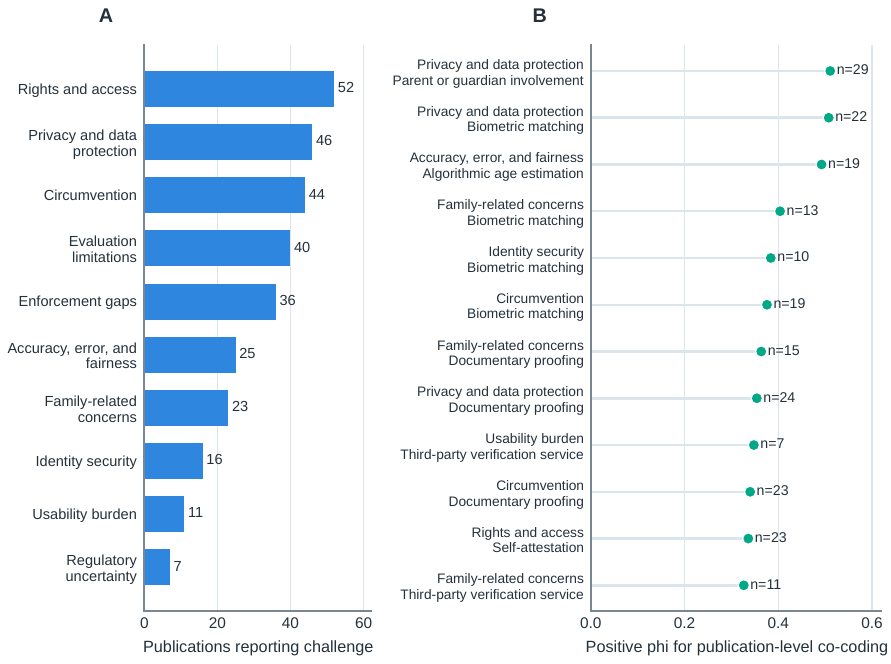}
  \caption{Reported challenge categories (A) and selected publication-level challenge--component associations (B). Panel B shows the 12 largest positive associations meeting the support criteria, after omitting one pair whose two categories were assigned from the same family-related wording (Appendix~\ref{app:analysis}); n gives the number of publications carrying both codes. Counts and coefficients are descriptive.}
  \Description{Two panels. Panel A is a bar chart of challenge category frequencies: rights and access 52, privacy and data protection 46, circumvention 44, evaluation limitations 40, enforcement gaps 36, accuracy, error, and fairness 25, family-related concerns 23, identity security 16, usability burden 11, and regulatory uncertainty 7. Panel B is a dot plot of the 12 largest positive phi coefficients for challenge and component co-coding, led by privacy with parent or guardian involvement at 0.511, privacy with biometric matching at 0.508, and accuracy, error, and fairness with algorithmic age estimation at 0.492, each annotated with the number of publications carrying both codes.}
  \label{fig:rq3-challenges-actors}
\end{figure*}

\findingclaim{Rights and access leads the ten reported challenge categories, while accuracy, error, and fairness ranks sixth.} Rights and access is the most frequent challenge category at 61\% (N=52), followed by privacy and data protection at 54\% (N=46), circumvention at 52\% (N=44), evaluation limitations at 47\% (N=40), enforcement gaps at 42\% (N=36), and accuracy, error, and fairness at 29\% (N=25). Family-related concerns, identity security, usability burden, and regulatory uncertainty occur less often (Figure~\ref{fig:rq3-challenges-actors}A).

\findingclaim{The most frequent challenges describe different points of failure, from unequal access burdens to unenforced rules.} Under rights and access, children's rights research examines expression, inclusion, and the unequal burden of proving one's age \cite{livingstone2024children,stardust2024mandatory}. Privacy research focuses on data collection, surveillance, secondary use, and data minimization \cite{scheffler2024systems,beltrn2024implications}. Circumvention research examines how users bypass checks through spoofing, replay attacks, false documents, and borrowed identities \cite{vanderhof2022we}; in one evaluation, replaying photographs to age estimators left them unable to tell the attacks from bona fide images \cite{korshunov2024vulnerability}. Research on evaluation limitations describes the difficulty of testing age-assurance systems, including incomplete datasets and weak external validation. Finally, enforcement research asks whether platforms and sellers implement the required checks; compliance audits of online sellers repeatedly find checks that rest on easily bypassed self-reports or are absent altogether \cite{egan2023absence,terala2023access,reile2025alcohol}.

\findingclaim{Privacy pairs most strongly with parent or guardian involvement and biometric matching, whereas accuracy, error, and fairness pairs with algorithmic age estimation.} These associations use the publication-level component indicators defined in Appendix~\ref{app:codebook}, a set of 12 that is wider than the four evaluating components named in RQ2 and that adds, among others, parent or guardian involvement, biometric matching, and third-party verification services. Privacy co-occurs with parent or guardian involvement at \(\phi=0.511\) and with biometric matching at \(\phi=0.508\), while accuracy, error, and fairness co-occur with algorithmic age estimation at \(\phi=0.492\) (Figure~\ref{fig:rq3-challenges-actors}B). Because these indicators are coded per publication, their counts differ from the process-level counts in RQ2.

\subsubsection{Explicit Research Gaps Center on Evaluation and Implementation}

\findingclaim{Among the gaps that prior work identifies as unresolved, evaluation limitations are the most frequent, followed by enforcement gaps.} Evaluation limitations appear as an unresolved gap in 45\% (N=38) of publications and enforcement gaps in 20\% (N=17). Rights and access ties with accuracy, error, and fairness at 18\% (N=15) each, and privacy appears in 15\% (N=13). Regulatory uncertainty, circumvention, family-related concerns, identity security, and usability burden occur less often. What these gaps call for differs by mode of inquiry. System evaluations call for larger or representative datasets and testing beyond benchmark settings \cite{hossain2023case,west2024picture}. Compliance audits call for continued monitoring of implementation \cite{egan2023absence,reile2025alcohol}, while legal and privacy analyses ask how to minimize identification yet still produce an actionable age-related claim \cite{scheffler2024systems,livingstone2024children}.

\subsubsection{Operational Roles Span Multiple Actors, but Accountability Links Remain Unspecified}
\label{sec:rq3-actors}

Who could answer for the reported challenges depends on which actors prior literature places in operational roles, and seven types recur. Platforms or service operators appear most often at 66\% (N=56), followed by governments or regulators at 60\% (N=51), parents, guardians, or families at 49\% (N=42), and users at 45\% (N=38). Verification service providers and retail or delivery actors each appear in 36\% (N=31) of publications, and technical developers or systems least often. Platforms configure age-assurance processes, providers evaluate signals, families authorize access, and regulators define or interpret obligations \cite{livingstone2024children}. Retail or delivery actors are expected to check eligibility at purchase or delivery \cite{reile2025alcohol}, and users supply the evidence---often a date of birth, or a parent's email address to which the platform sends a consent request \cite{vanderhof2022we}.

\findingclaim{The review literature fails to link actors in operational roles to false rejections, sensitive data exposure, exclusion, circumvention, or remediation.} Nor do the actor counts isolate minors. Because the ``user'' category in prior work often merges minors, adults, and generic consumers, these counts cannot reveal how often publications assign minors a separate operational role. The absence of links between the reported challenges and the operational roles is the accountability gap.

\begin{tcolorbox}[enhanced, colback=cognitiveblue!6, colframe=cognitiveblue!6, boxrule=0pt, arc=4pt, fuzzy shadow={2pt}{-2pt}{0pt}{0.2pt}{black!15}]
\textbf{\textcolor{cognitiveblue}{Key Takeaway:}} 
While prior work prioritizes rights, access, and privacy, it rarely connects system failures to the institutional actors operating the process. This systemic disconnect leaves an accountability gap, where no named actor is tied to a wrongfully excluded user or to a clear route for remediation.
\end{tcolorbox}

\section{Discussion}
\label{sec:discussion}

Age assurance became a policy expectation before it became a settled technical category, and the 85 reviewed publications span that gap. Communities that share few methods share one term, ``age verification,'' and each uses it for the part its methods reach \cite{reile2025alcohol,anda2020assessing,murray2025cyber}. Beneath the shared term lies one sequence, from what a user hands over to what the service does with it, and no community names that sequence as a whole. We develop that sequence into a unit of comparison, use it to identify what the reviewed publications leave unresolved, and close with the research those gaps require.

\subsection{The Age-Assurance Process Framework}
\label{sec:taxonomy}
Age assurance reached our SLR from several research contexts, each with its own question. Most of the reviewed publications appeared in public health or computing research, and a smaller group in governance research (Section~\ref{sec:rq1}). Public health research asks whether a retailer refuses a sale \cite{egan2023absence,williams2020sales,wang2025e}. Computing research asks how closely a model estimates an age from a face or a gesture \cite{west2024picture,bao2023general,hossain2023case}. Governance research asks what a statute requires of a service and what that service owes a regulator \cite{grimmelmann2024return,persson2024as,adrogu2026playground}. All three report that a service ``verifies age,'' so two publications can share the term while studying different operations. An online dispensary that admits a visitor on a click-through confirmation \cite{terala2023access} and an adult platform that inspects a government-issued identity document \cite{murray2025cyber} both ``verify age'' under this usage. A shared term is supposed to make results comparable; here it does the opposite.

\definecolor{aamut}{HTML}{55565A}
\definecolor{aaedg}{HTML}{B9BBBE}
\definecolor{aagry}{HTML}{80828A}
\definecolor{aaacc}{HTML}{B4560C}   \definecolor{aaacct}{HTML}{FBEADB}
\definecolor{aa1}{HTML}{2F6F9E}   \definecolor{aa1t}{HTML}{EDF3F8}
\definecolor{aa2}{HTML}{6A5AA6}   \definecolor{aa2t}{HTML}{F1EFF7}
\definecolor{aa3}{HTML}{1F7A6B}   \definecolor{aa3t}{HTML}{E8F3F0}
\definecolor{aa4}{HTML}{55636E}   \definecolor{aa4t}{HTML}{F0F2F4}

\begin{figure*}[t]
\centering
\resizebox{\textwidth}{!}{%
\begin{tikzpicture}[
  x=1cm,y=1cm,
  every node/.style={inner sep=0pt,outer sep=0pt},
  aahd/.style={font=\small\bfseries,text=white,anchor=center},
  aaref/.style={font=\footnotesize,text=aamut,anchor=center},
  aant/.style={font=\small,anchor=center,align=center},
  aaac/.style={font=\small\bfseries,text=aaacc,anchor=center,align=center},
  aae/.style={draw=aaedg,line width=0.6pt,-{Stealth[length=4pt,width=2.8pt]}},
  aaea/.style={draw=aaacc,line width=1.0pt,-{Stealth[length=4.6pt,width=3.2pt]}},
  aalg/.style={font=\footnotesize,text=aamut,anchor=north,align=center,
               execute at begin node={\hyphenpenalty=10000\exhyphenpenalty=10000\relax}},
  aagap/.style={font=\scriptsize,text=aagry,anchor=north,align=center,
                execute at begin node={\hyphenpenalty=10000\exhyphenpenalty=10000\relax}},
  aalgd/.style={font=\footnotesize,text=aamut,anchor=west},
]

\fill[aa1,rounded corners=3pt] (0.15,-0.52) rectangle (3.87,0.0);
\node[aahd] at (2.01,-0.26) {Input evidence or signal};
\node[aaref] at (2.01,-0.78) {\S\ref{sec:rq2-processes}};
\fill[aa2,rounded corners=3pt] (4.53,-0.52) rectangle (8.25,0.0);
\node[aahd] at (6.39,-0.26) {Evaluating component};
\node[aaref] at (6.39,-0.78) {\S\ref{sec:rq2-processes}};
\fill[aa3,rounded corners=3pt] (8.91,-0.52) rectangle (12.63,0.0);
\node[aahd] at (10.77,-0.26) {Age-related claim};
\node[aaref] at (10.77,-0.78) {\S\ref{sec:rq2-rep}, \S\ref{sec:rq2-decisions}};
\fill[aa4,rounded corners=3pt] (13.29,-0.52) rectangle (17.01,0.0);
\node[aahd] at (15.15,-0.26) {Downstream decision};
\node[aaref] at (15.15,-0.78) {\S\ref{sec:rq2-decisions}};

\draw[aaea] (3.87,-1.25) to[out=0,in=180,looseness=0.9] (4.50,-1.25);
\draw[aae] (3.87,-2.03) to[out=0,in=180,looseness=0.9] (4.50,-2.03);
\draw[aaea] (3.87,-2.81) to[out=0,in=180,looseness=0.9] (4.50,-2.81);
\draw[aae] (3.87,-3.59) to[out=0,in=180,looseness=0.9] (4.50,-3.59);
\draw[aaea] (8.25,-1.25) to[out=0,in=180,looseness=0.9] (8.88,-3.20);
\draw[aaea] (8.25,-2.81) to[out=0,in=180,looseness=0.9] (8.88,-3.20);
\draw[aae] (8.25,-3.59) to[out=0,in=180,looseness=0.9] (8.88,-3.20);
\draw[aae] (8.25,-2.03) to[out=0,in=180,looseness=0.9] (8.88,-3.20);
\draw[aae] (8.25,-2.03) to[out=0,in=180,looseness=0.9] (8.88,-1.25);
\draw[aae] (8.25,-2.03) to[out=0,in=180,looseness=0.9] (8.88,-2.03);
\draw[aaea] (12.63,-3.20) to[out=0,in=180,looseness=0.9] (13.26,-2.03);
\draw[aaea] (12.63,-3.20) to[out=0,in=180,looseness=0.9] (13.26,-3.20);
\draw[aaea] (12.63,-3.20) to[out=0,in=180,looseness=0.9] (13.26,-3.98);
\draw[aae] (12.63,-1.25) to[out=0,in=180,looseness=0.9] (13.26,-2.03);
\draw[aae] (12.63,-2.03) to[out=0,in=180,looseness=0.9] (13.26,-2.03);

\fill[white,draw=aaacc,line width=0.9pt,rounded corners=3pt] (0.15,-1.55) rectangle (3.87,-0.95);
\node[aaac,text width=3.62cm] at (2.01,-1.25) {Documentary identity data};
\fill[white,draw=aa1,line width=0.6pt,rounded corners=3pt] (0.15,-2.33) rectangle (3.87,-1.73);
\node[aant,text width=3.62cm] at (2.01,-2.03) {Facial imagery};
\fill[white,draw=aaacc,line width=0.9pt,rounded corners=3pt] (0.15,-3.11) rectangle (3.87,-2.51);
\node[aaac,text width=3.62cm] at (2.01,-2.81) {Self-reported age};
\fill[white,draw=aa1,line width=0.6pt,rounded corners=3pt] (0.15,-3.89) rectangle (3.87,-3.29);
\node[aant,text width=3.62cm] at (2.01,-3.59) {Platform metadata};
\fill[white,draw=aaacc,line width=0.9pt,rounded corners=3pt] (4.53,-1.55) rectangle (8.25,-0.95);
\node[aaac,text width=3.62cm] at (6.39,-1.25) {Documentary proofing};
\fill[white,draw=aa2,line width=0.6pt,rounded corners=3pt] (4.53,-2.33) rectangle (8.25,-1.73);
\node[aant,text width=3.62cm] at (6.39,-2.03) {Algorithmic age estimation};
\fill[white,draw=aaacc,line width=0.9pt,rounded corners=3pt] (4.53,-3.11) rectangle (8.25,-2.51);
\node[aaac,text width=3.62cm] at (6.39,-2.81) {Self-attestation};
\fill[white,draw=aa2,line width=0.6pt,rounded corners=3pt] (4.53,-3.89) rectangle (8.25,-3.29);
\node[aant,text width=3.62cm] at (6.39,-3.59) {Platform account gate};
\fill[white,draw=aa3,line width=0.6pt,rounded corners=3pt] (8.91,-1.55) rectangle (12.63,-0.95);
\node[aant,text width=3.62cm] at (10.77,-1.25) {Numeric age estimate};
\fill[white,draw=aa3,line width=0.6pt,rounded corners=3pt] (8.91,-2.33) rectangle (12.63,-1.73);
\node[aant,text width=3.62cm] at (10.77,-2.03) {Age-band estimate};
\fill[white,draw=aaacc,line width=0.9pt,rounded corners=3pt] (8.91,-3.50) rectangle (12.63,-2.90);
\node[aaac,text width=3.62cm] at (10.77,-3.20) {Threshold eligibility};
\fill[white,draw=aa4,line width=0.6pt,rounded corners=3pt] (13.29,-2.33) rectangle (17.01,-1.73);
\node[aant,text width=3.62cm] at (15.15,-2.03) {Digital access};
\fill[white,draw=aa4,line width=0.6pt,rounded corners=3pt] (13.29,-3.50) rectangle (17.01,-2.90);
\node[aant,text width=3.62cm] at (15.15,-3.20) {Regulated transaction};
\fill[white,draw=aa4,line width=0.6pt,rounded corners=3pt] (13.29,-4.28) rectangle (17.01,-3.68);
\node[aant,text width=3.62cm] at (15.15,-3.98) {Account participation};

\fill[white,draw=aaacc,line width=0.9pt,rounded corners=1.6pt] (0.90,-4.90) rectangle (1.16,-4.70);
\draw[aaea] (1.28,-4.80) -- (1.83,-4.80);
\node[aalgd] at (1.99,-4.80) {Orange traces two processes that share no evidence and reach the same claim};
\draw[aae] (10.91,-4.80) -- (11.46,-4.80);
\node[aalgd] at (11.62,-4.80) {Grey marks the other links the review records};
\end{tikzpicture}}
\caption{The age-assurance process. Each column represents one role that a mechanism label leaves unstated, and the nodes and edges are the categories and links the reviewed publications record most often. The second column holds verification and estimation together, which is why age assurance serves as an umbrella rather than a synonym for age verification (Sections~\ref{sec:umbrella} and~\ref{sec:rq2-term}). Orange traces two processes that share no input evidence yet reach the same age-related claim, and that claim then informs decisions whose errors cost different things. Furthermore, the extracted records connect none of these system failures to an actor who answers for them or to the remedies owed to excluded users (Section~\ref{sec:rq3-actors}). The figure reports which links appear, not how often.}
\Description{A four-column diagram with lines joining the columns. The columns are the four roles of an age-assurance process: input evidence or signal, evaluating component, age-related claim, and downstream decision, each heading in a filled color bar with the Findings section beneath it. On the left, four inputs -- documentary identity data, facial imagery, self-reported age, and platform metadata -- each connect to the component that acts on them: documentary proofing, algorithmic age estimation, self-attestation, and a platform account gate. Four of those components converge on a single node, threshold eligibility, drawn in a contrasting accent color, and lines then fan out from it to three decisions: digital access, a regulated transaction, and account participation. Algorithmic age estimation additionally reaches a numeric age estimate and an age-band estimate. The accent traces two complete paths, from documentary identity data and from self-reported age, into threshold eligibility and out again, giving the figure an hourglass shape. All boxes are white with a colored border. A legend states that orange traces two processes that share no evidence and reach the same claim, and that grey marks the other links the review records.}
\label{fig:arrangement}
\end{figure*}
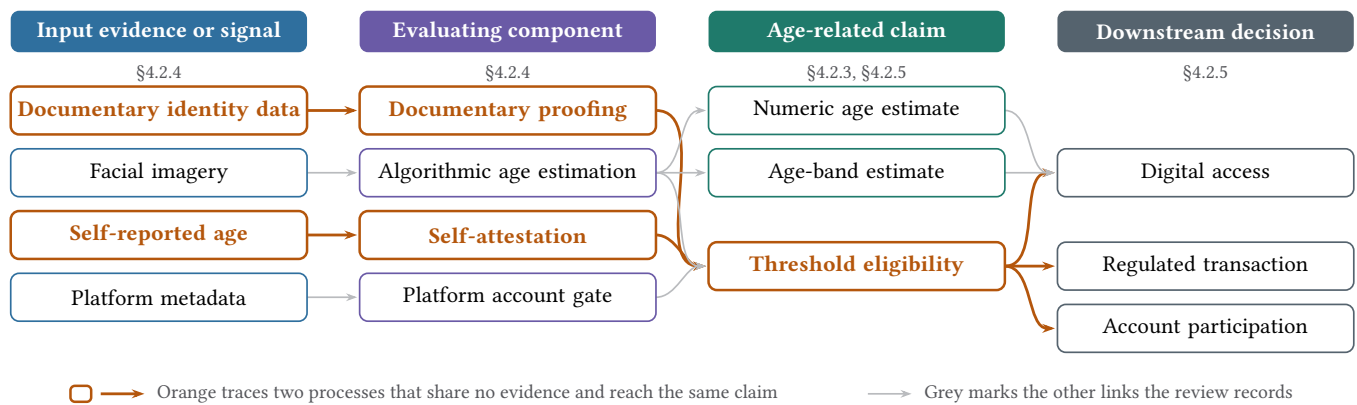

This divergence is more than terminological. Some reviewed publications report access control, others age estimation or prediction, and others identity or evidence validation, and many describe more than one at once (Section~\ref{sec:rq2-term}). A mechanism label specifies neither the evidence a check accepts nor the claim it establishes, so the same evaluating component connects different input evidence to different claims from one publication to the next \cite{association2024ieee,hossain2023case,vanderhof2022we}.

The \emph{age-assurance process} we defined in Section~\ref{sec:sociotechnical} is the unit that records what a mechanism label leaves open. Figure~\ref{fig:arrangement} shows its four roles with the categories the publications record most often. 
Comparability requires more than the four roles. A publication must also report how the claim errs near the threshold, what a wrongful refusal or admission costs, what becomes of the evidence afterward, and what route remains to contest a refusal. These are reporting requirements rather than scoring criteria. We leave the process unscored so that a documentary check and a facial estimator can be compared without ranking one above the other \cite{murray2025cyber,west2024picture}. The question thus shifts from ``which mechanism is strongest'' to ``which complete process did a publication study.''



At the \emph{input} stage, an age-assurance process makes its first demand of a user. Its three most frequent inputs---documentary identity data, facial imagery, and self-reported age---impose different costs on that user (Section~\ref{sec:rq2-processes}). A document check requires an identifier that outlives the transaction, and the reviewed publications often raise concerns about identity exposure and about the exclusion of users without conventional documents \cite{scheffler2024systems,stardust2024mandatory}. A facial check demands biometric data and raises concerns about demographic disparity and spoofing \cite{west2024picture,korshunov2024vulnerability}. A self-report asks for almost nothing and is as easily defeated as it is offered \cite{egan2023absence,vanderhof2022we}. The input stage therefore sets the privacy burden and the friction a user meets before any component has evaluated anything.

At the \emph{claim} stage, these differences narrow and provenance, the record of which input and component produced a claim, disappears. A government document and a checkbox both most often support threshold eligibility (Appendix~\ref{app:supplementary} lists the publications recording each), whereas facial imagery more often supports an estimate. The claim that reaches the service does not reveal which of the two produced it (Section~\ref{sec:rq2-rep}). A claim that a user is ``over eighteen'' therefore hides the privacy cost of producing it, which attaches to the evidence a check accepts rather than to the claim itself \cite{beltrn2024implications}. It also hides how the evaluating component fails, because an estimator's error near a threshold behaves differently from a documentary check that matches or fails. Cryptographic credential specifications already allow a claim to travel with a secure record of its source \cite{w3c2025,sdjwt2025,iso2021mdl}, yet such credentials remain among the least reported inputs in our corpus.

At the \emph{decision} stage, threshold eligibility informs digital access most often, but also regulated transactions and account participation (Section~\ref{sec:rq2-decisions}). One form of claim therefore serves decisions whose errors carry unequal costs. A wrongful refusal costs a moment at a website in one case and denies a lawful purchase in another. A wrongful admission yields a personalized feed in one and access to a regulated product in another. Regulation answers this asymmetry with \emph{proportionality}, the principle that the required certainty should track the cost of an error, but the instruments that state it differ. The UK Children's Code ties the required certainty to the risks a service creates for children \cite{ico2020aadc}, whereas Ofcom applies a single ``highly effective'' age-assurance standard to every service its duties cover \cite{ofcom2025hea}. Neither states the unit on which required certainty and error cost are compared, and the age-assurance process supplies one. Read against a proportionality requirement, the process implies a reverse order of design reasoning. The downstream decision determines the claim a service needs, the claim narrows the evidence a check may accept, and the evaluating component a mechanism label describes is chosen last.

Of the four roles, the evaluating component is the easiest for a provider to swap out, and compliance audits repeatedly find the least demanding one, self-attestation, in place (Section~\ref{sec:rq3}). What a provider must install instead is a legal question. Legal analysis dominates the second-largest topical domain, legal and regulatory frameworks, and it often interprets statutory obligations without measuring deployed outcomes \cite{pasquale2022consent,persson2024as}. This body of work therefore says much about what a service must do and little about what those actions establish. A duty attached to a ``mechanism'' inherits that gap, because the label states neither the claim a check produces nor the decision it informs. Recent law shows what that permits. The US Supreme Court upheld a state age-verification statute for material obscene to minors, while leaving mandates on fully protected speech unsettled \cite{fsc2025}. Subsequent statutes still differ on what counts as a compliant check \cite{liu2026tailoring}. One enforcement instrument promises forbearance if age data is minimized and deleted promptly \cite{ftc2026b}, but it attaches these conditions to the evidence without defining the claim the decision requires. A duty written against the complete process would define the required claim together with the admissible evidence. Because this SLR evaluates no statute's validity, it offers regulators that structural unit for policy design rather than a legal verdict.

The nearest test of this framework is the age signal that operating systems and app stores are adopting \cite{apple2026dar,google2026agesignals} under recent state statutes \cite{ca2025ab1043,tx2025sb2420}. Considered as an age-assurance process, such a signal does more than move the check from the service to the platform: it fixes the age-related claim before the downstream decision occurs. Its input evidence still ranges from self-reports to payment or facial checks, so the variation this framework records at the input stage remains. What the statutes standardize is the claim, fixing one set of age bands for each covered application and confining a developer's use of the signal to compliance purposes \cite{ca2025ab1043,tx2025sb2420}. Because the claim is fixed upstream of distinct downstream decisions, these statutes remove the point at which proportionality could be applied, before anyone has measured what a fixed claim costs. Neither statute requires the signal to carry its provenance \cite{ca2025ab1043,tx2025sb2420}, yet platform interfaces already report the evidence source alongside the age band \cite{apple2026dar,google2026agesignals}. Conditioning consequential decisions on that provenance would restore proportionality, separating a band that rests on verified or guardian-set evidence from one resting on a declared birth date alone.

This age-assurance framework also reframes the question the field is asked most often. Whether ``age verification works'' depends on which part of the process a study measures, and work assessing these technologies argues that effectiveness cannot be reported in isolation \cite{lueks2026}. Each community measures only the role its methods can see, and our corpus records no evaluation spanning all four roles, even where a publication specifies all four. This partial measurement falls hardest on minors. Age-group information identifies a population far more often than it structures a measurement (Section~\ref{sec:rq2-role}), and the ``user'' actor category that might have isolated minors often merges them with adults. A review of systems built to protect minors therefore cannot report in full what those systems do to minors. Two of the four roles carry those effects for minors. For research on children's privacy, the evidence a check demands is where data protection becomes a design variable \cite{crepax2022information,stoilova2020}. For research on children's safety, the downstream decision is where unjust exclusion becomes a measurable outcome rather than an accepted cost \cite{uncrc2021digital,wisniewski2024moving}. Young people already argue on these terms, objecting to what a check demands and to the access it removes rather than to the accuracy of the evaluating component \cite{woodley2025australian}. Meeting them there requires treating age assurance as a complete sociotechnical process rather than an inventory of isolated mechanisms.

\subsection{Sociotechnical Gaps in Age-Assurance Processes}
\label{sec:gaps}

Section~\ref{sec:sociotechnical} defined an \emph{age-assurance process} as one complete sequence spanning all four roles, from the evidence a user supplies or a system observes to the decision that evidence licenses. Mapping the reviewed publications onto these four roles exposes four sociotechnical gaps in age assurance. \textbf{First, an age-assurance process is described by what it collects and not by what it keeps.} While our extraction gave privacy and data protection an evidence field of its own, with retention among the concerns it covers, the reviewed publications frame this as a theoretical concern rather than a measurement. Privacy is the second most frequently reported challenge, recurring alongside components that read biometric data or involve parents (Section~\ref{sec:rq3}). Yet, what a service does with the evidence after the decision is almost never stated: a system evaluation of facial age estimation reports how accurately a model reads a face and stops there \cite{anda2020assessing,bao2023general}. One publication follows camera frames down to the on-device file where a platform writes its age predictions \cite{west2024picture}, but that depth of tracing is rare in our corpus. Two processes can therefore be compared on what they demand of a user but not on what they retain, even though retention is the half of the privacy cost that outlives the check.

\textbf{Second, components are evaluated in aggregate and enforced at a threshold, and the reviewed publications do not connect the two.} Although publications state the age thresholds their processes enforce (Section~\ref{sec:rq2-role}), evaluation limitations remain the gap they acknowledge most often. System evaluations frequently ask for larger datasets and testing beyond benchmark settings \cite{hossain2023case,west2024picture}, yet neither request reaches the specific threshold a decision turns on. Aggregate accuracy hides subgroup error \cite{buolamwini2018gender}, and estimator error varies across the adolescent age range \cite{hanaoka2024fate}; thus, an aggregate figure says little about the exact ages where a threshold falls. A parent or a regulator asks first how often a system is wrong about a fifteen-year-old, and no reviewed publication answers that question.

\textbf{Third, age-related decisions lack systemic accountability.} The reviewed publications name actors from the platform configuring the process to the regulator writing the obligation, and they describe what a wrong decision costs. Our extracted records do not connect the two, so no named operational roles are tied to a wrongfully excluded user or to a remedy (Section~\ref{sec:rq3-actors}). Naming a role is not the same as establishing accountability, which requires a forum in which someone owes an explanation and consequences follow \cite{bovens2007analysing}. Two things such a forum needs are missing here. First, a claim that arrives without its provenance gives an excluded user no grounds on which to contest a refusal, and research on algorithmic explanations finds no format that reliably reads as just \cite{binns2018reducing}. Second, someone must answer for a system failure, and assigning responsibility grows harder when a process spans organizations that can deflect blame \cite{nissenbaum1996accountability,wieringa2020account}. Recording provenance is a solvable engineering problem \cite{w3c2025,sdjwt2025}; constituting the accountability forum is not. The candidates who might answer are also becoming fewer, as web-scale verification concentrates in a few providers \cite{minocha2026papers}. We consider this accountability gap as the most important open problem in age-assurance research.

\textbf{Fourth, age-assurance processes treat children as isolated users and ignore the household.} The access, transaction, and account decisions in our corpus all fall on one individual user, yet the reviewed publications record parents taking part in producing the claim that decision rests on. Parent or guardian involvement is its own evaluating component category, and parent-related evidence is a recorded input (Section~\ref{sec:rq2-processes}). Work outside this literature shows what registering only the individual leaves out. Parents often help children bypass age checks, treating that help as parental mediation rather than as a security breach \cite{boyd2011why,livingstone2008}. Families negotiate device rules at home rather than having rules imposed on one member \cite{luria2025teen,hiniker2016dinner}. Evidence requests also presume a privacy competence children do not have, because children reason about privacy in interpersonal terms and overlook institutional data collection \cite{stoilova2020,zhao2019}.

This mismatch between the household and the individual explains why a security-centered view of youth circumvention fails. An age check that asks a shopper to click a box is defeated the moment anyone clicks it \cite{terala2023access,rhee2025gating}. The technical answer is a stronger check, such as continuous facial estimation, but that escalation extracts a heavier privacy cost from the children it aims to protect \cite{west2024picture,scheffler2024systems}. The stronger check is then defeated in its turn, by replayed samples or by structural loopholes \cite{korshunov2024vulnerability,eltaher2025loophole}, and the escalation resumes. Two research lenses divide this cycle between them, and each explains only its own half. A security lens treats circumvention as an adversary to defeat, which warrants the escalation but leaves the privacy cost unaccounted. A childhood-studies lens treats circumvention as user resistance, which registers that cost but sets aside the need for the check. Neither lens accounts for the parents who help their children bypass these checks at home. Seeing that help requires studying children's circumvention with children, rather than inferring it from server logs \cite{nansen2026phreaking}.

The first two gaps, retention and threshold error, have technical answers. Cryptographic credentials can carry a secure record of provenance and disclose one attribute at a time \cite{w3c2025,sdjwt2025,iso2021mdl}. Disclosing one attribute at a time limits what a service needs to keep, and the provenance record supplies what an excluded user needs to contest a decision. Similarly, a reporting convention can break out error rates by age band. Privacy-preserving credentials and structured trials already exist to build on \cite{ec2025avblueprint,accs2025trial}. The other two gaps, accountability and the household, cannot be fixed with software. A credential cannot force a platform to offer an appeal, and a reporting standard still leaves a system evaluating individuals rather than households. Treating a social problem as a settled engineering question is techno-legal solutionism \cite{reid2025nerd}. Calling a process ``solved'' because it carries provenance and reports error bands, while accountability and the household remain open, is exactly that. Cryptography cannot resolve social accountability \cite{bellovin2025privacy}. Two duties therefore remain. First, an institution must owe the wrongfully excluded user an explanation and a route to appeal. This duty is in tension with enforcement forbearance requiring prompt deletion of age evidence \cite{ftc2026b}. Keeping a durable record of the claim's provenance, rather than the raw evidence, resolves this tension, allowing an institution to answer for a refusal without holding the data it is required to delete. Second, a system must recognize the household that produces the age-related claim rather than the isolated individual it currently registers.

\begin{table*}[tb]
\caption{Four directions for research on the age-assurance process, one per row. Read a row left to right: the gap this SLR records and the sections that report it, the three measures a study closing that gap would report, and the prior work it can build on. Entries are not matched item by item across columns. Rows are ordered by the community equipped to run the work, not by importance.}
\label{tab:future-work}
\scriptsize
\begin{tabularx}{\textwidth}{P{0.13\textwidth}P{0.22\textwidth}P{0.24\textwidth}Y}
\toprule
\textbf{Research direction} & \textbf{The gap, and where this SLR reports it} & \textbf{What a study would report} & \textbf{What already exists to build on} \\
\midrule

\textbf{Study what an age check does to the user it refuses.}\newline \emph{HCI}
& Neither a compliance audit nor a system evaluation records the user's interaction with the check (Section~\ref{sec:rq1}).\newline
Actors are named without being connected to a refused user (Section~\ref{sec:rq3-actors}).
& (1)~Who was refused and what the refusal cost them.\newline
(2)~What alternative pathway existed and how long recovery took.\newline
(3)~Whether the user could use that pathway without an adult.
& Ceremony and human-in-the-loop analysis of the process around a check \cite{ellison2007ceremony,cranor2008human}.\newline
Procedural-justice outcomes for an adverse decision \cite{binns2018reducing}.\newline
A study design that reaches children directly \cite{nansen2026phreaking}.\newline
An automated audit that records where age checks fail without reaching the user \cite{figueira2026actions}. \\
\addlinespace

\textbf{Measure outcomes for adolescents, not for a generic population.}\newline \emph{Children's online privacy and safety}
& Age-group information rarely structures a measurement (Section~\ref{sec:rq2-role}).\newline
The category that holds users merges minors with everyone else (Section~\ref{sec:rq3-actors}).
& (1)~Error and refusal broken out by age band.\newline
(2)~Recovery of access for young people recruited as participants.\newline
(3)~The family treated as the site that produces an age claim.
& Participatory work with children \cite{druin2002}.\newline
Parental mediation as the frame for a jointly produced claim \cite{livingstone2008,boyd2011why}.\newline
Rights and resilience framings that admit unflattering outcomes \cite{uncrc2021digital,livingstone2025there,wisniewski2024moving}.\newline
Population-scale outcome measurement \cite{barnes2026assessing}. \\
\addlinespace

\textbf{Evaluate a complete age-assurance process, not a single component.}\newline \emph{Usable security and privacy}
& The recorded processes preserve what a process connects without reporting how it performs at any role (Section~\ref{sec:rq2-processes}).\newline
Evaluation limitations are among the challenges raised most often (Section~\ref{sec:rq3}).
& (1)~Disclosure per check and error across age bands.\newline
(2)~What follows a refusal and what the process costs an attacker.\newline
(3)~Whether the people subject to it accept the result.
& Selective disclosure specified and deployed in mobile credentials \cite{w3c2025,sdjwt2025,iso2021mdl}.\newline
A white-label application and a national trial \cite{ec2025avblueprint,accs2025trial}.\newline
A comparison framework in which no scheme dominates on every criterion \cite{bonneau2012quest}.\newline
A proposal to assess effectiveness alongside side effects and acceptance \cite{lueks2026}. \\
\addlinespace

\textbf{Specify what a process must report before an obligation can be audited.}\newline \emph{Policy, governance, and standards research}
& Legal analysis interprets an obligation without measuring what it produces once deployed (Section~\ref{sec:taxonomy}).\newline
Named actors are not tied to a failure or a remedy (Section~\ref{sec:rq3-actors}).
& (1)~The claim an obligation requires, the evidence admissible for it, and the decision it licenses.\newline
(2)~Error at the enforced threshold reported by age band.\newline
(3)~What a refused user is told, and by whom.
& A standard that separates the policy maker, the provider, and the relying party \cite{iso2025ageassurance}.\newline
A single scale of assurance built from stated indicators \cite{association2024ieee}.\newline
A national trial treating layered methods as the route to a risk-appropriate decision, while finding no single solution effective in every deployment \cite{accs2025trial}.\newline
A reference implementation to build against \cite{ec2025avblueprint}. \\

\bottomrule
\end{tabularx}
\end{table*}

\subsection{Future Work for Age-Assurance Research}
\label{sec:future}
What prior work reports as a challenge and what it proposes as future work diverge. ``Rights and access'' and ``privacy'' lead the challenges the reviewed publications report, yet only a minority of those publications state them as unresolved gaps. ``Evaluation limitations'' shows the opposite pattern and leads the future-work agendas (Section~\ref{sec:rq3}). The divergence follows from methodological limits rather than neglect. A legal analysis that identifies an access harm lacks the tools to measure it \cite{grimmelmann2024return,stardust2024mandatory}, and a compliance audit is not designed to ask what a refusal costs the user \cite{egan2023absence,reile2025alcohol}. Our framework instead treats privacy exposure and unjust exclusion as properties of the complete age-assurance process rather than of any one role. Table~\ref{tab:future-work} sets out four directions on that unit, each assigned to the community equipped to run it.

\textbf{Study what an age check does to the user it refuses.} A refusal is the moment a user learns what was decided about them and whether they can appeal it. The compliance audits and system evaluations that dominate our corpus stop at the system \cite{terala2023access,rhee2025gating,anda2020assessing,korshunov2022face}, and no publication we reviewed brings HCI's own instruments to the refused user.

\textbf{Measure outcomes for adolescents, not for a generic population.} Two commitments make this harder than it looks. A study must hold protection and participation together, because a wrongly refused child loses a right rather than a convenience \cite{uncrc2021digital}. It must also report outcomes that do not flatter the intervention, which fields organized around protecting youth may find uncomfortable \cite{wisniewski2024moving}. Neither commitment is new, but neither has been applied to age assurance outside one population-scale study \cite{barnes2026assessing}.

\textbf{Evaluate a complete age-assurance process, not a single component.} Evaluating one process end to end runs into ownership: the input, the component, the claim, and the decision sit with different parties, so the evaluation needs access that no single party can grant. The work is nonetheless urgent. The operating-system age signal discussed in Section~\ref{sec:taxonomy} is already shipping, and it fixes the claim before any party has evaluated the process that claim belongs to.

\textbf{Specify what a process must report before an obligation can be audited.} Policies attach their requirements to different parts of the process: Australia regulates the outcome \cite{au2024smma}, Ofcom the overall effectiveness \cite{ofcom2025hea}, California the age bands \cite{ca2025ab1043}, and Texas the party running the check \cite{tx2025sb2420}. An auditor therefore has no common quantity to check. Even where regulation ties required certainty to data-processing risks \cite{ico2020aadc}, none of these instruments requires error at the enforced threshold to be reported. Nor does the international standard fix the thresholds a claim must meet \cite{iso2025ageassurance}.

\section{Limitations}
\label{sec:limitations}

Our review is bounded by its English-language, 2020--2026 search, the selected databases, available full texts, and age-related queries. The retained search records also lack execution dates, source-specific syntax, and deduplication rules. This SLR therefore characterizes the retrieved publications rather than claiming exhaustive coverage. Our Discussion also cites work our search did not retrieve, including publications that appeared while the corpus was being screened; that work informs our interpretation but sits outside the 85 publications and every count we report.

Our coding procedure bounds what the reported figures demonstrate. A single researcher performed primary relevance screening, an LLM performed the data extraction, and the deterministic coding rules were developed with AI coding assistance. Validation covered every extracted entry, but because the two human coders audited disjoint halves of the corpus, the $\kappa$ reported in Section~\ref{sec:validation} measures agreement between the LLM extraction and the adjudicated human record, not between coders. Re-running the coding rules reproduces their assignments by construction, so that check confirms consistent application rather than correct categorization, and no coder recoded the assignments independently. The text pipeline also did not inspect image-only figures and tables. Source excerpts, locations, raw outputs, and fixed coding procedures make our categorization decisions traceable, but a coded category may not record every distinction a publication draws.

The synthesis describes the prevalence of research attention rather than the strength of empirical evidence, because it applies no quality appraisal for individual study designs. Publication-level co-occurrences identify thematic associations rather than causal relations. The age-assurance process records are excerpt-supported interpretations, and because contexts, actors, and challenges were coded separately, no result we report attributes a system failure to a named actor.

\section{Conclusion}
Across the 85 publications, ``age verification'' is not a single mechanism. The publications describe processes that differ in their input evidence or signals, evaluating components, age-related claims, and downstream decisions. Those differences bear on the privacy a process demands of a user and on what a wrongful refusal costs that user. By connecting these four roles, the \textit{Age-Assurance Process Framework} lets HCI research compare processes structurally rather than treat evaluating components as interchangeable. By exposing an accountability gap, it directs research to the missing connections between institutional actors, system failures, and the remedies owed to excluded children.

\bibliographystyle{ACM-Reference-Format}
\bibliography{references}

\appendix

\section{Search and Screening Procedure}
\label{app:search}

Section~\ref{sec:search} summarizes the search and screening; this appendix reports the full query, the retrieval sequence, the yields by source, and every exclusion category.

The three Boolean blocks, joined by AND, are:

\begin{quote}
\small
\sloppy
\texttt{("age verification" OR "age gating" OR "age assurance" OR "age estimation" OR "age check*")}\\
\textbf{AND} \texttt{("minor*" OR "child*" OR "teen*" OR "adolescen*" OR "youth")}\\
\textbf{AND} \texttt{("digital" OR "social media" OR "online" OR "internet" OR "platform*" OR "chatbot*" OR "AI")}
\end{quote}

We began developing the search in early February 2026. We exported and reviewed initial results from the ACM Digital Library, IEEE Xplore, and USENIX in the third week of the month, then extended the strategy to the remaining sources. We then built the 2020--2026 range into the retrieval criteria and reran every search. Retrieval is therefore an iterative February 2026 search rather than a single execution date. The project record preserves the common query, the refinement sequence, and the source yields, but not a final interface-specific query log for every source.

The final searches returned 707 records: 177 from the ACM Digital Library, 47 from IEEE Xplore, 306 from Scopus, 173 from Web of Science, and four from USENIX or exploratory Google Scholar.

Duplicate removal preceded relevance coding and removed 268 duplicates. The ledger begins with the deduplicated set and does not preserve the software, matching fields, or cluster-resolution notes used in this step.

Sequential relevance coding began in late February 2026. The primary screener applied five exclusion criteria in order: prior systematic reviews (15 records), medical or non-digital age assessment (181), work without a substantive connection to minors (73), child-sexual-abuse-material detection or age estimation (20), and criminal-justice age estimation outside the digital-access boundary (three). Together they produced 292 criterion-based exclusions. Final eligibility decisions excluded another 56 records: 46 adjacent or otherwise out-of-scope records, three annotated summaries, one non-English full text, and six records with no documented reason. These categories are mutually exclusive and sum to 348.

Two records had inconsistent screening indicators, and neither preserved criterion-level adjudication notes. We therefore reconciled both against the 85 publications included in extraction. One carried an inclusion marker together with a child-sexual-abuse-material exclusion code and was classified as excluded because it was absent from the included publications. A poster carried a nonstandard inclusion marker and was classified as retained because it had retrievable full text and was included in extraction. The reconciled selection contains 91 retained reports, six unretrieved reports, and 85 publications in the synthesis. This reconciliation was post hoc, not a prospectively documented adjudication. Figure~\ref{fig:prisma-flow} reports the final flow. The review was not preregistered.

\section{Finding Themes and Their Supporting Publications}
\label{app:themes}

Table~\ref{tab:themes} maps the coded categories whose counts the Findings report to the publications recorded under them, so that a reader can move from a reported count to the literature behind it. Processes whose source left the downstream use unspecified have no row, because they name no decision category; Appendix~\ref{app:codebook} reports how many there are. Categories within a family are non-exclusive unless the Findings state otherwise, so one publication can appear in several rows, and the rows together cite all 85 included publications.
\clearpage
\onecolumn
\begingroup
\tiny
\setlength{\tabcolsep}{3pt}
\begin{longtable}{@{}>{\raggedright\arraybackslash\leftskip=1em\relax}p{0.25\textwidth}P{0.028\textwidth}P{0.65\textwidth}@{}}
\caption{Finding themes and the included publications recorded under each. Counts are unique publications, and the families follow the order in which the Findings report them. Cross-tabulations, the study-context categories, the publication-level component indicators, and the component counts that apply the generic-wording rule appear in the Findings rather than here.}
\label{tab:themes}\\
\toprule
\hspace*{-1em}\textbf{Theme} & \textbf{N} & \textbf{Included publications} \\
\midrule
\endfirsthead
\toprule
\hspace*{-1em}\textbf{Theme} & \textbf{N} & \textbf{Included publications} \\
\midrule
\endhead
\bottomrule
\endfoot
\multicolumn{3}{@{}p{0.93\textwidth}}{\textbf{RQ1: Research domains, methods, and policy settings} (Section~\ref{sec:rq1})} \\*[2pt]
\multicolumn{3}{@{}p{0.93\textwidth}}{\emph{Topical domain.} What each publication examines within age assurance. Domains are non-exclusive, so one publication can carry several.} \\*
Platform safety & 44 & \cite{west2024picture,gupta2024advancements,scheffler2024systems,persson2024as,odeigah2025underage,alam2022based,machdi2024prediction,shah2023applying,hilbert2025bigtech,poh2025biometric,nali2025blasting,jfatt2025debate,skinner2025debate,livingstone2024children,haluska2024concept,p2023content,murray2025cyber,kennedy2025demographic,khlerdauner2025child,pasquale2022consent,dobbs2025discreetshipping,gaiha2025e,vassey2023e,abdulla2025i,beltrn2024implications,adib2023improvising,wisniewski2024moving,jarvie2024government,reid2025nerd,woodley2025australian,lakis2023pose,nagata2025prevalence,yar2020protecting,cover2025restricting,obrien2020social,eltaher2025loophole,grimmelmann2024return,livingstone2025there,kong2024tobacco,hossain2020touch,tananjaya2023towards,vanderhof2022we,pulfrey2022zoom,abidin2025tiktok} \\
Legal and regulatory frameworks & 37 & \cite{scheffler2024systems,williams2020sales,persson2024as,reile2025alcohol,bearman2025analysis,anda2020assessing,poh2025biometric,nali2021characterizing,gupta2024child,mulligan2025children,livingstone2024children,colbert2020content,murray2025cyber,khlerdauner2025child,pasquale2022consent,dobbs2025discreetshipping,abdulla2025i,association2024ieee,beltrn2024implications,adib2023improvising,crepax2022information,donaldson2025monitoring,loparco2024retail,sneyd2024alcohol,reid2025nerd,ekambaranathan2022poster,nagata2025prevalence,yar2020protecting,cover2025restricting,adrogu2026playground,grimmelmann2024return,livingstone2025there,vanderhof2022we,bujalski2024alcohol,azagba2023loopholes,stardust2024mandatory,henes2025youth} \\
\end{longtable}
\endgroup

\twocolumn
\section{Evidence Fields and Extraction Procedure}
\label{app:fields}

\subsection{Field Definitions and Extraction Settings}

Section~\ref{sec:extraction} summarizes the extraction; this appendix reports the field definitions, the model settings, and the extraction prompt. Table~\ref{tab:field-families} defines the 17 evidence fields and one optional synthesis memo, and Appendix~\ref{app:prompt} reports the evidence contract that governs what each field may hold. Three entries, in two fields, contain several extracted items supported by one excerpt for the field as a whole rather than by a separate excerpt for every item; Appendix~\ref{app:analysis} reports these exceptions.

Extraction ran on 3 July 2026. Requests used GPT-5.4 with temperature 0, JSON-object response mode, a 12,000-token output limit for full-text and merge requests, an 8,000-token limit for chunk requests, a 360-second timeout, and up to three attempts. The pipeline used \texttt{pypdf} to process pages separately, collapsed repeated blank lines, and inserted explicit \texttt{[PAGE n]} boundaries.

Publications containing at most 70,000 extracted characters were processed in one request. Longer publications were divided at page boundaries into sections of approximately 60,000 characters. Each section produced evidence only from its pages; the final merge considered at most six non-missing candidates per field and could not introduce new evidence. Seventy publications used one full-text request, 12 were divided into two sections, one into three, and two into 11. We retained the machine-readable output from each request, including output repaired only to produce valid JSON, along with token counts, completion status, and processing mode. Adding record identifiers to 11 synthesis memos and three excerpts did not change their substantive content.

\begin{table*}[t]
\centering
\scriptsize
\caption{Seventeen evidence fields and one optional synthesis memo. Definitions were developed for this review; citations identify representative included publications.}
\label{tab:field-families}
\begin{tabular}{p{0.2\textwidth} p{0.55\textwidth} p{0.15\textwidth}}
\toprule
\textbf{Evidence field} & \textbf{Definition} & \textbf{Primary use} \\
\midrule
Age-group focus & Population labels, age thresholds, or age-based cohorts explicitly studied, protected, compared, or evaluated. Broad labels are not treated as equivalent developmental categories \cite{hossain2023case,schiff2021accessing}. & RQ2 \\
Topical domain & The substantive setting in which a publication discusses age assurance, such as platform safety or regulated-product access \cite{west2024picture,reile2025alcohol,hossain2023case}. & RQ1 \\
Mode of inquiry or evidence & The methodological approach a publication uses to substantiate its findings, such as system evaluation or evidence synthesis \cite{hossain2023case,reile2025alcohol,schiff2021accessing}. & RQ1 \\
Legal or policy context & Laws, regulations, standards, policy domains, or institutional obligations explicitly tied to age assurance \cite{persson2024as,livingstone2024children,association2024ieee}. & RQ1 \\
Exact terminology & Exact age-verification, estimation, assurance, gating, or related wording used by the publication \cite{vanderhof2022we,livingstone2024children,hossain2023case}. & RQ2 \\
Definition or operational description & Text, a figure, or a table that explains what age assurance or a related term means in the publication's own account \cite{livingstone2024children,association2024ieee,vanderhof2022we}. & RQ2 \\
Decision purpose & The publication-level objective or institutional function for which age-related information is produced or used \cite{persson2024as,egan2023absence,hossain2023case}. & RQ2 \\
Named operational actor or authority & Actors or authorities named as supplying evidence, operating, authorizing, overseeing, contesting, or remediating a process. This field does not assign normative accountability \cite{livingstone2024children,reile2025alcohol,woodley2025australian}. & RQ3 \\
Technical component & Algorithmic, hardware, or data-processing operations that estimate age, evaluate evidence, or apply an age-related rule \cite{hossain2023case,poh2025biometric,association2024ieee}. & RQ1--RQ3 \\
Social, human, or institutional component & An action a person or organization performs while collecting evidence, reviewing it, authorizing access, or overseeing an age-assurance process \cite{schiff2021accessing,reile2025alcohol,livingstone2024children}. & RQ1--RQ3 \\
Age-related attribute or status established & The attribute or status a process seeks to establish. In the analysis, threshold eligibility, numeric-age estimates, age-band estimates, and age-verification status are treated as age-related claims; authorization and relationship outputs are retained separately \cite{hossain2023case,west2024picture,association2024ieee}. & RQ2 \\
Input evidence, signal, or predictor feature & The evidence, observable signal, or predictor feature an evaluating component uses, such as documentary identity data or facial imagery \cite{west2024picture,hossain2023case,association2024ieee}. & RQ2 \\
Direct input--component--claim--decision linkage & A numbered process record that stores the input evidence or signal, evaluating component, age-related claim or other reported output, and downstream decision together with a designated source excerpt \cite{hossain2023case,schiff2021accessing,poh2025biometric}. & RQ2 \\
Privacy and data protection & Privacy and data-protection concerns identified by the publication, such as surveillance or retention \cite{scheffler2024systems,beltrn2024implications,livingstone2024children}. & RQ3 \\
Consequences of incorrect decisions & Consequences of false acceptance, false rejection, misclassification, exclusion, or other erroneous operation \cite{west2024picture,abdulla2025i,stardust2024mandatory}. & RQ3 \\
Circumvention or resilience & Bypass and robustness concerns, such as spoofing or borrowed evidence \cite{korshunov2024vulnerability,schiff2021accessing,vanderhof2022we}. & RQ3 \\
Author-stated remaining challenges & Limitations, future-work needs, unresolved tensions, or research gaps explicitly stated by the publication \cite{hossain2023case,west2024picture,reile2025alcohol}. & RQ3 \\
Synthesis memo & A short note that connects evidence across fields. It may support interpretation but is not counted as a publication-level evidence field. & Interpretation only \\
\bottomrule
\end{tabular}
\end{table*}

\subsection{Prompt Design for Evidence Extraction}
\label{app:prompt}

Listing~\ref{lst:extraction-prompt} summarizes the extraction instruction issued for every publication. It presents the extraction task, the evidence contract, the source hierarchy and field boundaries, the labeling rules that tie each value to its own excerpt, and the output schema.

The listing gives the field keys as issued and keeps the four-question numbering under which the prompt was run. Under that numbering, conceptualization and operationalization were separate questions (RQ2 and RQ3) and challenges were RQ4; this paper merges the first two as RQ2 and reports challenges as RQ3. Several field keys carry mechanism wording that this review reports under its own component and claim vocabulary. Table~\ref{tab:field-families} lists the same fields in that vocabulary, in the same order, and against the research question each one serves.

\begin{lstlisting}[style=codingprompt,caption={Summary of the extraction instruction issued for each publication.},label={lst:extraction-prompt}]
Task:
Read one screened publication and return structured, auditable extraction notes for a systematic review of digital age verification, age assurance, age estimation, age prediction, age gating, and age-based access controls for minors. Use only the supplied publication text, and do not rely on outside knowledge. Do not assign relevance tiers or exclusion decisions, because the corpus has already been screened. Do not produce final qualitative codes.

Evidence Contract:
1. Every analytic field returns exactly four subfields: value, supporting_excerpt, page_or_section, and reasoning.
2. supporting_excerpt is copied verbatim from the supplied text. Completeness and traceability outrank brevity, so an excerpt may run to several paragraphs when a manual audit needs that much. Normalizing whitespace is acceptable; paraphrasing is not, and OCR artifacts and line-break hyphenation must not be silently repaired in a way that changes the visible wording.
3. value may be normalized or summarized for coding and spreadsheet review, except where a field definition requires the publication's own wording. If a value normalizes author wording, reasoning briefly explains the normalization.
4. terminology_used_exact is the main exception, and its value preserves author wording.
5. If the text does not explicitly support a field, return "None" for all four subfields. Do not infer a value, backfill from a neighboring publication, or use outside knowledge.
6. Every substantive value item must be supportable from the same field's supporting_excerpt. This is semantic support, so a normalized value word need not appear verbatim. If an item is not supportable, add the quotation that supports it, narrow the value, remove the item, or return None for the field.
7. Keep author-stated remaining challenges separate from the synthesis memo, and do not place a model-inferred gap in paper_stated_remaining_challenges_or_open_questions.
8. reasoning is a short evidence-based rationale, not hidden chain-of-thought.
9. Return strict JSON only, with every key present in the order given. Do not add provenance, alignment, duplicate-quote, nested-entry, or human-audit-status subfields.

Evidence Source Hierarchy:
1. Use the whole publication, but prioritize its own contribution sections. Do not record a background literature-review claim as this publication's mechanism, finding, or challenge unless the authors adopt it as their own framing, proposal, result, or conclusion.
2. RQ1 and RQ2 fields may draw on the abstract, introduction, background, and literature review.
3. For RQ3 fields, prioritize the sections that describe what the publication itself studies, proposes, evaluates, or recommends: Methods, Materials, Study Design, System Design, Model or Algorithm, Dataset, Measures, Results, Findings, Analysis, Proposed Approach, Framework, or Standard Requirements.
4. For RQ4 fields, prioritize Discussion, Limitations, Conclusion, Future Work, Policy Implications, Recommendations, or Threat Model.
5. Introduction and background evidence may supplement core-section evidence, but it must not substitute for it.
6. Where no empirical Methods section exists, use the closest equivalent core section, such as a legal analysis, a conceptual framework, a technical proposal, a specification, or a set of recommendations.
7. If introduction or literature-review text supports a substantive mechanism or challenge field, reasoning must briefly explain why that excerpt reflects the authors' own framing, proposal, or conclusion rather than a summary of prior work.

Missing Evidence:
1. Not every publication contains evidence for every field. This is expected in a corpus mixing technical, legal, policy, empirical, review, and conceptual work.
2. Do not accept weak proxy evidence to avoid None. A facial age-estimation study may not discuss legal context, and a policy analysis may not specify data inputs.
3. None is analytically meaningful. A None value may indicate that the publication does not address that dimension rather than that the extraction failed, and later synthesis may use None values to identify absent or underdeveloped dimensions.

Field Boundaries:
1. age_related_attribute_or_status_established holds the target construct, meaning what the process seeks to know, classify, verify, compare, or decide: exact age, age range, age bin, minor or adult status, under-13 status, over-18 status, claimed-age match, maturity category, or eligibility for restricted access.
2. data_inputs_signals_or_predictor_features holds the evidence used to establish that attribute: government identification, self-declared date of birth, selfie, camera frame, voice sample, touch gesture, account metadata, shipping or payment data, classifier score, or intermediate model output.
3. If a term could fit either field, code it by its function in the publication. An age bin being predicted is an attribute; a child score used to infer whether a user is a child is a signal, unless the publication treats the score as the final status.
4. A mechanism field describes the process, an attribute field describes what the process establishes, and an input field describes the evidence the process uses.
5. Keep technical mechanisms separate from social, human, and institutional ones.
6. Do not add a general conceptualization field. Use age_verification_definition, stated_purpose_of_age_verification, and responsible_actor_or_authority instead.

Item-Level Evidence Mapping:
1. Where a value holds several distinct items, label them [Item 1], [Item 2], and label the matching evidence [Item 1 evidence] or [Items 1-3 evidence] inside the same field's supporting_excerpt.
2. Several excerpts may support one item, labeled [Item 1 evidence A] and [Item 1 evidence B].
3. Labels are required for any field holding two or more items and optional for single-item fields. Do not pair a labeled value with an unlabeled excerpt.
4. If a value item is normalized from author wording, include the author wording in the matching supporting_excerpt and explain the normalization in reasoning.
5. Before returning JSON, check field by field that a matching label in the same field's supporting_excerpt covers every [Item n] and every [Link n], and repair any label that is missing.

Precise-Item Self-Check:
1. A precise item in a value requires its exact wording, number, label, table or figure text, caption, or close OCR equivalent inside that same field's supporting_excerpt.
2. Precise items include numeric thresholds such as under 13 or over 18; explicit age bins such as 0-2, 3-9, 10-19; named laws, standards, policies, and jurisdictions; named model outputs such as baby score or child score; named documents such as government ID or passport; figure, table, and axis labels; and author-defined terms.
3. Where an excerpt supports only a broader idea, use the broader value. An excerpt naming an age range does not support an age-bin value unless a further quotation in the same field supplies the age-bin wording.
4. If a precise item appears elsewhere in the publication, add that quotation to the field rather than relying on support recorded in another field.

Linkage Rule:
1. attribute_signal_linkages records explicit or closely local relations among process, input or signal, age-related attribute or status, and decision or use.
2. Format each relation as [Link n] mechanism/process -> input/signal/predictor feature -> age-related attribute/status -> decision/use.
3. Components may be normalized but must stay defensible from the same field's supporting_excerpt, labeled by link and by component where the parts appear in separate passages.
4. If the publication states that government identification, date of birth, facial age estimation, parental consent, payment or shipping data, or any other signal is used to determine an age threshold such as under 13 or over 18, extract that relationship as a linkage.
5. Do not create a link because an input and an attribute both appear somewhere in the publication. The publication must connect them inside one mechanism, method, system description, analysis, requirement, policy rule, or closely adjacent explanation.
6. If the publication establishes an age status without stating what follows from it, write "decision/use not specified". Do not supply a plausible decision.
7. Where no defensible relation can be established, return None for the field.

Key Order:
Return the seventeen evidence fields in this order: age_group_focus, domain_or_topic_area, study_design_or_evidence_type, legal_or_policy_context, terminology_used_exact, age_verification_definition, stated_purpose_of_age_verification, responsible_actor_or_authority, technical_mechanisms, social_human_or_institutional_mechanisms, age_related_attribute_or_status_established, data_inputs_signals_or_predictor_features, attribute_signal_linkages, privacy_considerations, unintended_consequences_if_wrong, circumvention_or_resilience_challenges, and paper_stated_remaining_challenges_or_open_questions.
optional_synthesis_memo follows them. It is a reviewer note whose text must state that it awaits cross-publication validation, not a publication-stated finding. uncertainty_notes closes the object as a free-text note outside the field set.

Output:
Return one JSON object holding paper_title, the seventeen evidence fields and optional_synthesis_memo in the order listed above, and uncertainty_notes. paper_title and uncertainty_notes are plain strings, and every other key takes the object shape below.
{
  "paper_title": "<exact title, PDF filename, or None>",
  "<field_name>": {
    "value": "<concise extracted value, item-labeled where needed, or None>",
    "supporting_excerpt": "<verbatim excerpt with matching item labels, or None>",
    "page_or_section": "<page, section heading, table or figure label, or None>",
    "reasoning": "<short evidence-based rationale, or None>"
  },
  "uncertainty_notes": "<ambiguity, weak evidence, inaccessible sections, or None>"
}
\end{lstlisting}

\section{Coding Rules and Component Definitions}
\label{app:codebook}

\subsection{Category Construction and the Component Codebook}

Section~\ref{sec:coding} describes the 111 coding rules; this appendix reports the rule groups (Table~\ref{tab:normalization-rules}), the component definitions, and the resulting coding coverage.

\begin{table*}[H]
\caption{Category groups and how each is assigned: the evidence each group draws on, how a rule assigns it, and what the resulting label supports. Every group except the generic-component group comes from the 111 coding rules; that one applies the separate rule its own row states.}
\label{tab:normalization-rules}
\scriptsize
\begin{tabularx}{\textwidth}{P{0.20\textwidth}P{0.27\textwidth}YY}
\toprule
\textbf{Category group} & \textbf{Source evidence} & \textbf{Assignment procedure} & \textbf{What the label supports} \\
\midrule
Topical domain and mode of inquiry & Topical-domain and mode-of-inquiry evidence & Match predefined terms without regard to capitalization; count repeated matches within a publication once. & Non-exclusive description of the included publications rather than a formal classification of the research area. \\
Terminology, definition, age-group function, and purpose & Corresponding evidence fields & Assign every matching terminology or meaning category. & Categories developed for this review; paired excerpts retain source wording. \\
Legal or policy context & Legal and policy evidence & Assign every matched policy. & Identifies discussed context, not legal applicability or effect. \\
Publication-level component, challenge, and actor & Component, challenge, and actor evidence & Assign every matching broader category and count repeated matches within a publication once. & Appearance in the same publication does not establish a direct link among categories. \\
Direct four-part process & Numbered process records & Separate the input evidence or signal, evaluating component, age-related claim or other reported output, and downstream decision; apply the first matching rule in each documented priority order. & Each process is traceable to a designated excerpt; the component category may use the rule described below when the source wording is generic. \\
Component category when wording is generic & Component wording plus its linked input or output & When the component description does not identify a category, the generic-wording rule assigns one from the linked input or output. & A category assigned this way is not established by the source wording itself, so the main component-process panels exclude it. \\
Study context & Title, venue, publication type, domain, and mode-of-inquiry wording & Assign one label through a fixed priority order. & Describes study context rather than disciplinary identity. \\
\bottomrule
\end{tabularx}
\end{table*}

Table~\ref{tab:component-codebook} defines the 13 component categories the rules assign. Twelve serve as non-exclusive publication-level indicators, and all 13 are available to the ordered rule that assigns a component within a four-part process. \emph{Rule-based access control} occurs only in process records, because no publication-level indicator corresponds to it. The labels describe what a component does rather than where it appears in a technical system. The codebook foregrounds six of these as reference categories. Four of them, documentary proofing, algorithmic age estimation, self-attestation, and the platform account gate, carry the component comparisons in Section~\ref{sec:rq2-processes}; the other two appear where the analysis requires them.

\begin{table*}[t]
\centering
\scriptsize
\caption{Component codebook. The first six categories are the reference categories the Findings use when comparing components, selected for frequency in the excerpt-supported processes and for analytic centrality; the remaining seven are retained rather than merged, because collapsing them would attribute an operation to a component that the source did not describe. Definitions were developed for this review, and the citations in each definition identify representative included publications. The typical recorded process summarizes the input, claim, and decision most often recorded under that category across all 251 processes, and the coding boundary states what the category does not establish.}
\label{tab:component-codebook}
\begin{tabular}{@{}p{0.135\textwidth}p{0.275\textwidth}p{0.225\textwidth}p{0.275\textwidth}@{}}
\toprule
\textbf{Component category} & \textbf{Definition} & \textbf{Typical recorded process} & \textbf{Coding boundary} \\
\midrule
Documentary proofing & Evaluates documentary identity or age evidence issued or accepted by an institution. \cite{association2024ieee,crepax2022information,schiff2021accessing} & Document input linked to threshold eligibility or age-verification status. & The document is the input; documentary proofing is the component that evaluates it. \\
Algorithmic age estimation & Infers numeric age, an age band, or threshold likelihood from a biometric or behavioral signal. \cite{hossain2023case,west2024picture,durgam2024estimation} & Facial, vocal, touch, or behavioral input linked to an estimated claim. & The model is the component; facial, vocal, touch, and behavioral data are inputs. \\
Self-attestation & Accepts a user-provided age or date of birth without independent documentary corroboration in the described step. \cite{vanderhof2022we,persson2024as,woodley2025australian} & Self-report linked most often to threshold eligibility. & A later independent check is coded separately rather than folded into self-attestation. \\
Platform account gate & Applies account metadata, a stored platform age state, or a platform rule to participation or access. \cite{odeigah2025underage,eltaher2025loophole,kong2024tobacco} & Platform metadata linked to threshold eligibility and to account participation. & An account gate is an operational component and does not reveal how the age state was first produced. \\
Biometric matching & Compares a biometric sample with reference evidence to corroborate identity or an associated age claim. \cite{poh2025biometric,eltaher2025loophole,vanderhof2022we} & Facial or other biometric input linked most often to an age-verification status. & Matching is distinct from estimating age directly from a biometric signal. \\
Transaction-based verification & Embeds an age-assurance process in a retail, payment, shipping, delivery, or in-person transaction procedure. \cite{egan2023absence,reile2025alcohol,schiff2021accessing} & Commerce or documentary evidence linked to a regulated-transaction decision. & The transaction context is central; documentary or human checks inside it may also receive their own component labels. \\
Human inspection & A person acting for the service, such as a clerk, moderator, or delivery driver, reviews evidence or observes the user as part of an access or transaction decision. \cite{schiff2021accessing,nali2021characterizing} & Human observation or documentary input linked to threshold eligibility in a regulated transaction. & Human observation and documentary identity data are inputs; the act of inspection is the component. \\
Third-party verification service & An intermediary provider performs, validates, or transmits an age check separately from the relying platform, retailer, or service. \cite{williams2020sales,association2024ieee,beltrn2024implications} & Documentary or facial input linked to threshold eligibility for digital access. & Naming a provider does not specify the evidence method or the age-related claim produced. \\
Privacy-preserving credential & Uses a cryptographic token, anonymous credential, or selective proof to establish an age predicate while limiting identity disclosure. \cite{scheffler2024systems,murray2025cyber,poh2025biometric} & Cryptographic-credential or documentary input linked to threshold eligibility for digital access or account participation. & The category describes a component design and does not by itself establish data minimization in deployment. \\
Parent or guardian involvement & A parent, guardian, or family member supplies consent evidence, authorization, or another operational action within the process. \cite{pasquale2022consent,abdulla2025i,vanderhof2022we} & Facial, parent-supplied, platform, or documentary input linked to an authorization or relationship status. & Operational involvement does not establish that a parent or guardian is responsible for a failure or a remedy. \\
Institutional enforcement & A law, regulator, standard, certification, policy, or organizational process requires, audits, or enforces an age-assurance obligation. \cite{hilbert2025bigtech,mulligan2025children,kong2024tobacco} & Platform-metadata or residual input linked to threshold eligibility in an access or transaction decision. & A policy mention is insufficient unless the excerpt describes an operational enforcement role. \\
Non-facial biometric estimation & Infers age from voice, speech, touch, gesture, behavior, or another non-facial biometric or behavioral signal. \cite{ilyas2020biometricaccessfilter,hossain2020touch,pulfrey2022zoom} & Non-facial biometric input linked to a numeric age estimate. & A mode-specific subset of estimation, retained so that non-facial signals remain visible rather than absorbed into facial estimation. \\
Rule-based access control & Applies explicit rule logic or a threshold test to permit, block, filter, warn, or otherwise govern access. \cite{alam2022based,haluska2024concept,p2023content} & A prior age or verification output linked to threshold eligibility for digital access. & Process records only. No publication-level indicator corresponds to it, and it is not merged with the account gate category. \\
\bottomrule
\end{tabular}
\end{table*}

\subsection{Coding Coverage and Publication Years}

Across categories assigned by ordered rules, 136 values initially matched more than one rule and were assigned by the documented priority order. Fifty-eight values could not be assigned to a named category and remained in residual categories, and 51 decision records remained unspecified. At least one named decision category was available for 79 of the 85 publications. Outputs such as authorization, relationships, consistency, or identity corroboration were retained but not counted as age-related claims. Because the categories have different denominators, these figures describe coding coverage rather than proportions or the accuracy of the category definitions.

Temporal analysis uses the publication year recorded during screening. For 14 publications, that year differs from the publisher online-first metadata, mainly because online-first and issue years differ.

\section{Association Measures and Extraction Validation}
\label{app:analysis}

Section~\ref{sec:synthesis} reports the association analysis and the validation result; this appendix gives the phi formula, the pairs meeting the support rule, and the derivation behind each validation figure.

For binary publication-level indicators \(X\) and \(Y\), phi is calculated as:

\begin{equation}
\phi =
\frac{n_{11}n_{00}-n_{10}n_{01}}
{\sqrt{(n_{11}+n_{10})(n_{01}+n_{00})(n_{11}+n_{01})(n_{10}+n_{00})}}.
\label{eq:phi}
\end{equation}

Here, \(n_{11}\) is the number of publications coded for both \(X\) and \(Y\), \(n_{10}\) is the number coded only for \(X\), \(n_{01}\) is the number coded only for \(Y\), and \(n_{00}\) is the number coded for neither category.

Of the 120 challenge--component pairs, 58 positive pairs met the support rule: the two categories appeared together in at least five publications, and each appeared in at least eight. Twenty negative pairs met it as well, the largest being enforcement gaps with algorithmic age estimation at $\phi=-0.39$. Figure~\ref{fig:rq3-challenges-actors}, Panel B, displays the 12 largest positive coefficients after omitting the family-related-concern and parent-or-guardian-involvement pair; the same family-related wording helped assign both categories and could inflate their association.

Of the 1,445 possible publication-by-field entries defined by the 17 evidence fields, 1,350 contain a value and a supporting excerpt; the remaining 95 contain neither. The optional synthesis memo contains a value and excerpt for all 85 publications but is not treated as an evidence field. All 251 four-part processes have correspondingly numbered excerpts. Two definition records and one terminology record each use one excerpt to support a multi-item field rather than linking every item to a separate excerpt. Independent human validation covered all 1,445 evidence-field entries, including these records and the ten quotation discrepancies flagged before validation, and Section~\ref{sec:validation} reports the result. After adjudication, we regenerated all reported frequencies, cross-tabulations, process counts, actor and challenge counts, and phi coefficients from the validated data.

Table~\ref{tab:validation-2x2} yields the agreement and $\kappa$ reported in Section~\ref{sec:validation} and the precision and recall figures below. Observed agreement is $(1{,}344+71)/1{,}445 = 0.979$, and chance agreement is $(1{,}350/1{,}445)(1{,}368/1{,}445)+(95/1{,}445)(77/1{,}445) = 0.888$, so $\kappa = 0.81$ (95\% CI [0.75, 0.88]). This coefficient compares the extraction against the adjudicated human record, not the two coders against each other. Treating field population as an information-retrieval task yields 99.6\% precision, 98.2\% recall, and an F1 score of 98.9\%; a stricter measure that treats all revisions as errors yields 98.3\% precision, 97.0\% recall, and an F1 score of 97.6\%.

Every rate reported in Section~\ref{sec:validation} runs optimistic, because the coders judged the model's populated output instead of coding from a blank record. That order makes an omission harder to notice than a wrong value and places the recall figures at an upper bound.

\begin{table}[htbp]
\centering
\caption{Agreement between LLM-assisted extraction and the adjudicated human record on the prior decision the coders made for every entry, namely whether the evidence field held evidence at all. Rows give the extraction outcome and columns the adjudicated reference.}
\label{tab:validation-2x2}
\small
\begin{tabular}{@{}lccc@{}}
\toprule
& \multicolumn{2}{c}{\textbf{Adjudicated human record}} & \\
\cmidrule(lr){2-3}
\textbf{LLM-assisted extraction} & \textbf{Held evidence} & \textbf{Held no evidence} & \textbf{Total} \\
\midrule
Populated the field & 1,344 & 6 & 1,350 \\
Returned no value & 24 & 71 & 95 \\
\addlinespace
\textbf{Total} & 1,368 & 77 & 1,445 \\
\bottomrule
\end{tabular}

\vspace{5pt}
\begin{minipage}{0.72\linewidth}
\footnotesize \emph{Note.} Cells count publication-by-field entries, and the 17 evidence fields applied to 85 publications give 1,445 entries. The off-diagonal cells hold the 24 entries recorded as evidence missed and the six recorded as an unsupported value, which Table~\ref{tab:validation-fields} distributes across fields. The 17 entries whose value or excerpt was revised fall inside the top-left cell of 1,344, because this adjudication recorded only whether a field held evidence.
\end{minipage}
\end{table}

Table~\ref{tab:validation-fields} reports the adjudicated outcome for each evidence field. Its validation RQ column maps the validation workbook's field-to-RQ assignments onto this paper's three research questions; the workbook numbered conceptualization and operationalization separately. For several fields the map differs from the primary manuscript use given in Table~\ref{tab:field-families}, and that difference is an analytic reassignment, not a change to the extracted evidence.

\begin{table*}[htbp]
\centering
\caption{Adjudicated human-validation outcome for each evidence field. Every field was validated across all 85 publications, so each row covers 85 entries and each entry contributes exactly one outcome. Rows are ordered by retention, then by the field order of Table~\ref{tab:field-families}.}
\label{tab:validation-fields}
\footnotesize
\begin{tabular}{@{}lrrrrc@{}}
\toprule
& & \multicolumn{3}{c}{\textbf{Entries corrected at adjudication}} & \\
\cmidrule(lr){3-5}
\textbf{Evidence field}
  & \begin{tabular}[b]{@{}c@{}}\textbf{Retained}\\\textbf{(\%)}\end{tabular}
  & \begin{tabular}[b]{@{}c@{}}\textbf{Value or}\\\textbf{excerpt}\\\textbf{revised}\end{tabular}
  & \begin{tabular}[b]{@{}c@{}}\textbf{Evidence}\\\textbf{missed}\end{tabular}
  & \begin{tabular}[b]{@{}c@{}}\textbf{Value not}\\\textbf{supported}\end{tabular}
  & \begin{tabular}[b]{@{}c@{}}\textbf{Validation}\\\textbf{RQ}\end{tabular} \\
\midrule
Privacy and data protection & 90.6 & 0 & 8 & 0 & 3 \\
Circumvention or resilience & 91.8 & 0 & 6 & 1 & 3 \\
Definition or operational description & 92.9 & 0 & 6 & 0 & 2 \\
Social, human, or institutional component & 92.9 & 4 & 0 & 2 & 2 \\
Named operational actor or authority & 95.3 & 3 & 1 & 0 & 2, 3 \\
Input evidence, signal, or predictor feature & 95.3 & 3 & 0 & 1 & 2 \\
Technical component & 96.5 & 3 & 0 & 0 & 2 \\
Decision purpose & 97.6 & 0 & 2 & 0 & 2 \\
Age-related attribute or status established & 97.6 & 2 & 0 & 0 & 2 \\
Direct input--component--claim--decision linkage & 97.6 & 1 & 0 & 1 & 2 \\
Legal or policy context & 98.8 & 0 & 0 & 1 & 1, 2, 3 \\
Exact terminology & 98.8 & 1 & 0 & 0 & 2 \\
Consequences of incorrect decisions & 98.8 & 0 & 1 & 0 & 3 \\
Age-group focus & 100.0 & 0 & 0 & 0 & 1 \\
Topical domain & 100.0 & 0 & 0 & 0 & 1 \\
Mode of inquiry or evidence & 100.0 & 0 & 0 & 0 & 1 \\
Author-stated remaining challenges & 100.0 & 0 & 0 & 0 & 3 \\
\midrule
\textbf{All fields} & 96.7 & 17 & 24 & 6 & \\
\bottomrule
\end{tabular}

\vspace{5pt}
\begin{minipage}{0.95\linewidth}
\footnotesize \emph{Note.} Retention is the percentage of a row's 85 entries that the coders left uncorrected; the three correction columns are entry counts, and across fields they give the 47 corrections reported in Section~\ref{sec:validation}. The final row pools all 1,445 entries, for which the validation RQ map does not apply.
\end{minipage}
\end{table*}

\section{Age Operationalization and Decision Purpose}
\label{app:visualizations}

\begin{figure*}[htbp]
  \centering
  \includegraphics[width=0.5\textwidth]{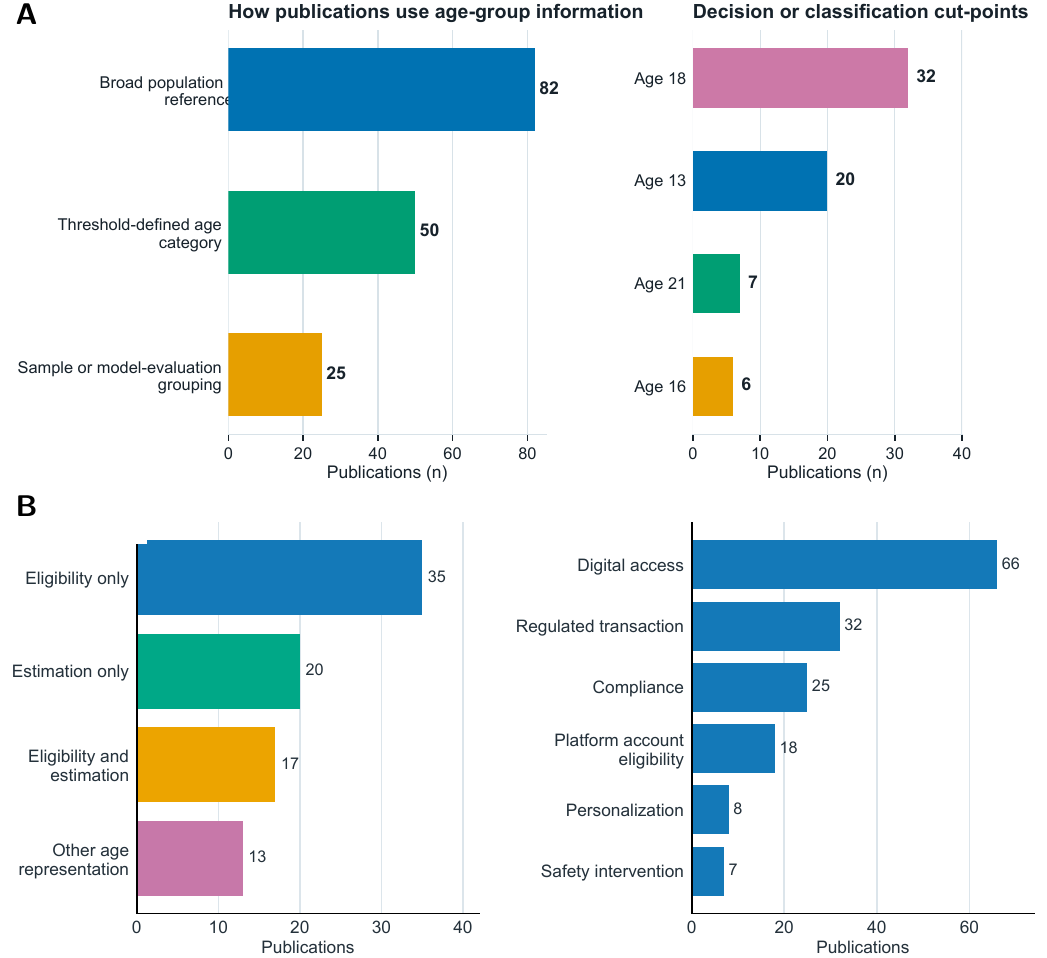}
  \caption{Age-group functions, cut-points, and age representation across the 85 publications. Panel A reports the functions age-group information serves, and the most frequent explicit cut-points. Panel B reports the four representation profiles, and the purposes for which publications use age information. Profiles are exclusive; functions, cut-points, and purposes are not.}
  \Description{Two vertically arranged panels. Panel A shows three functions for age-group information, broad population reference 82, threshold-defined age category 50, and sample or model-evaluation grouping 25, beside the most frequent explicit cut-points, age 18 with 32, age 13 with 20, age 21 with 7, and age 16 with 6. Panel B shows representation profiles, eligibility only 35, estimation only 20, both 17, and other age representation 13, beside decision purposes across all 85 publications, digital access 66, regulated transaction 32, compliance 25, platform-account eligibility 18, personalization 8, and safety intervention 7.}
  \label{fig:rq2-age-operationalization}
\end{figure*}

\begin{figure*}[htbp]
  \centering
  \includegraphics[width=0.5\textwidth]{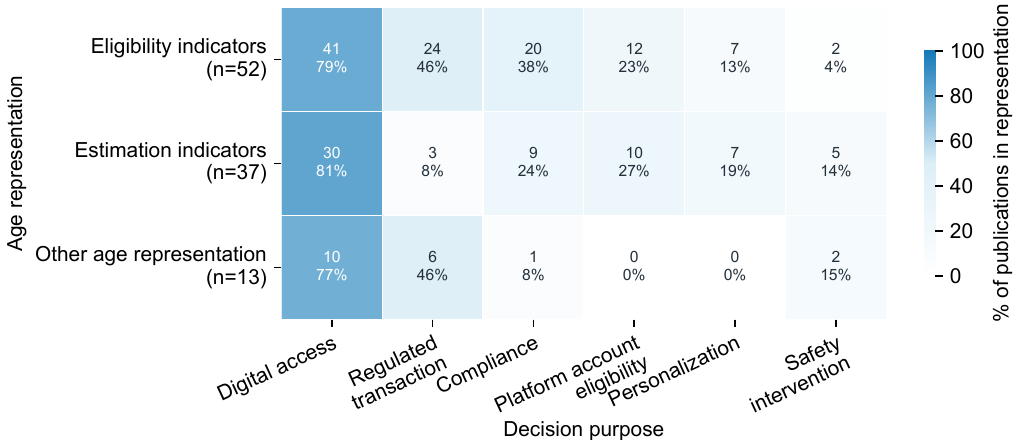}
  \caption{Publication-level co-coding between age representation and decision purpose. Cells count unique publications; percentages use the row total. The eligibility and estimation rows overlap, and decision purposes are non-exclusive, so a row can exceed 100\%.}
  \Description{A heatmap crossing three age-representation rows with six decision purposes; each cell gives a publication count and a row percentage. Eligibility indicators (n=52): digital access 41 (79\%), regulated transaction 24 (46\%), compliance 20 (38\%), platform-account eligibility 12 (23\%), personalization 7 (13\%), safety intervention 2 (4\%). Estimation indicators (n=37): digital access 30 (81\%), regulated transaction 3 (8\%), compliance 9 (24\%), platform-account eligibility 10 (27\%), personalization 7 (19\%), safety intervention 5 (14\%). Other age representation (n=13): digital access 10 (77\%), regulated transaction 6 (46\%), compliance 1 (8\%), platform-account eligibility 0, personalization 0, safety intervention 2 (15\%).}
  \label{fig:rq2-purpose}
\end{figure*}

The two figures below carry the publication-level detail behind Sections~\ref{sec:rq2-role} and~\ref{sec:rq2-rep}. Figure~\ref{fig:rq2-age-operationalization} reports the functions age-group information serves, the recurrent cut-points, the four representation profiles, and the purposes for which publications use age information. Figure~\ref{fig:rq2-purpose} crosses the two representations with those purposes and holds the counts behind the comparison in Section~\ref{sec:rq2-rep}.
\section{Recurrent Age-Assurance Processes}
\label{app:supplementary}

This appendix lists the recurrent age-assurance processes Section~\ref{sec:taxonomy} cites: combinations of the four roles that more than one publication records. Documentary proofing connected to documentary identity data and threshold eligibility appears in ten publications serving digital access \cite{abdulla2025i,crepax2022information,gaiha2025e,grimmelmann2024return,livingstone2025there,loparco2024retail,murray2025cyber,odeigah2025underage,reile2025alcohol,williams2020sales} and ten serving regulated transactions \cite{bertrand2025easy,colbert2020content,dobbs2025discreetshipping,gaiha2025e,hardie2022retail,henes2025youth,livingstone2024children,rhee2025gating,schiff2021accessing,sneyd2024alcohol}. Self-attestation connected to self-reported age and threshold eligibility appears in eight digital-access publications \cite{bertrand2025easy,hardie2022retail,livingstone2025there,odeigah2025underage,rhee2025gating,shroff2024marketing,terala2023access,woodley2025australian} and six regulated-transaction publications \cite{bujalski2024alcohol,colbert2020content,egan2023absence,nali2021characterizing,reile2025alcohol,riwubara2023they}. Each list enumerates every publication recording that process. These counts use all records, including some whose component category came from the generic-wording rule, so they show patterns under the review's coding rules rather than confirmation that each source used the label.


\end{document}